\documentstyle[a4,amsmath,amssymb,latexsym,epsfig,citesort]{article}
\def \d{{\mathrm{d}}}

\def \pd{\partial}

\def \tl#1{\overset{\kern 1pt\circ}{#1}}
\def \TL#1{\overset{\kern -3pt \circ}{#1}}
\def \TLL#1{\overset{\kern -7pt \circ}{#1}}

\begin{document}
\title{{\bf  On Dislocations in a Special Class of Generalized Elasticity}}

\author{{\bf Markus Lazar~$^{\text{a,b,}}$\footnote{Corresponding author.
{\it E-mail address:} lazar@lmm.jussieu.fr (M.~Lazar).}\;
, G{\'e}rard A.~Maugin~$^\text{a}$ and Elias C.~Aifantis~$^{\text{b,c}}$}\\ \\
${}^\text{a}$ Laboratoire de Mod{\'e}lisation en M{\'e}canique,\\
        Universit{\'e} Pierre et Marie Curie,\\
        4 Place Jussieu, Case 162,\\    
        F-75252 Paris Cedex 05, France\\
${}^\text{b}$ Laboratory of Mechanics and Materials, Polytechnic School,\\
Aristotle University of Thessaloniki, P.O. Box 468,\\
54124 Thessaloniki, Greece\\
${}^\text{c}$ Center for the Mechanics of Material Instabilities and Manufacturing Processes,\\
Michigan Technological University,\\
Houghton, MI 49931, USA
}

\date{\today}    
\maketitle
\begin{abstract}
In this paper we consider and compare 
special classes of static theories of gradient elasticity, nonlocal 
elasticity, gradient micropolar elasticity and nonlocal micropolar elasticity
with only one gradient coefficient.
Equilibrium equations are  discussed. The relationship between the gradient 
theory and the nonlocal theory is discussed for elasticity as well as
for micropolar elasticity. 
Nonsingular solutions for the elastic fields 
of screw and edge dislocations are given. 
Both the elastic deformation (distortion, strain, bend-twist) and the 
force and couple stress tensors do not possess any singularity
unlike `classical' theories. 
\end{abstract}
\vspace*{5mm}

\section{Introduction}
\setcounter{equation}{0}
The aim of this paper is to give an overview of a special class of 
generalizations of elasticity. 
In classical continuum 
mechanics, the elastic continuum is viewed as a collection of 
particles which have only three translational degrees of freedom.
The particles are taken without structure and are idealized as point
masses. 

The first attempts to modify the theory of elasticity were done 
by taking rotational degrees of freedom in addition to the translational ones.
In such media, which are called Cosserat (or micropolar) media,
every material particle is considered as a rigid volume element.
Each particle has six independent degrees of freedom: 
3 micro-rotational and 3 translational ones of the center.
Thus, a micropolar continuum can be described by geometrical points
to which a micro-rotation vector is attached. 
Accordingly, a particle is identified by its position vector and 
its micro-rotation vector. 

For the calculation of stresses, strains and distortions 
produced by defects (dislocations, disclinations, cracks)  
such theories break down near the defect lines because they are 
not valid at very small distances. 
For instance, the elastic stress field of a dislocation has a $1/r$-singularity
and the elastic couple stress field has a $1/r^2$-singularity.

Thus, generalized continuum theories which 
are able to explain the material behaviour on the nanoscale are needed.
Such theories can build bridges between continuum and atomic physics.
On the other hand, the singularities in the elastic fields of defects
should disappear in a theory which is also valid in the defect core region.   
The following question arises: What kind of generalized elasticity theories
are able to do that?

An extension of elasticity theory which goes into another direction 
than the Cosserat theory is 
the theory of nonlocal elasticity.
The concept of nonlocal elasticity was originally proposed 
by Kr{\"o}ner and Datta~\cite{KD66,Kroener67}, Edelen and Eringen~\cite{EE72}, 
Kunin~\cite{Kunin83} and some others.
Nonlocal theories were introduced to explain the material behaviour on
the nanoscale (e.g., in the core of defects).
Such a theory considers the inner structure of materials and
takes into account long-range (nonlocal) interactions.
In a continuum theory, nonlocality arises due to the finite 
range of interaction between material points. 
The long range of interaction implies that the quantities such as stress and energy
are functionals depending on the motion of all points of the body.
The nonlocal kernel, which appears in the nonlocal constitutive relation
between the stress and strain fields, weights the elastic constants of the material.
The kernel is supposed to model the details of the atomic interaction. 
The theory of nonlocal elasticity was successfully applied to
the calculation of the stresses produced by defects (cracks, dislocations, disclinations).
The main goal is the elimination of the classical unphysical singularities
of the stress 
fields~\cite{Eringen77a,Eringen77b,AE83,Eringen83,Eringen83b,Eringen02,Povstenko95,Povstenko95b,PM00,Lazar03a}. 
Thus the stress fields are smooth even in the core region.
In addition, nonlocal elasticity leads to a natural fracture criterion which is
based on the maximum stress hypothesis~\cite{Eringen83b,Eringen02}.
So, materials will fracture when the calculated maximum stress reaches the cohesive stress 
of atomic bonds.
However, the strain field singularities are still present in nonlocal elasticity.
\begin{figure}[t]\unitlength1cm
\centerline{
\begin{picture}(14,3)
\put(0.0,0.0){\epsfig{file=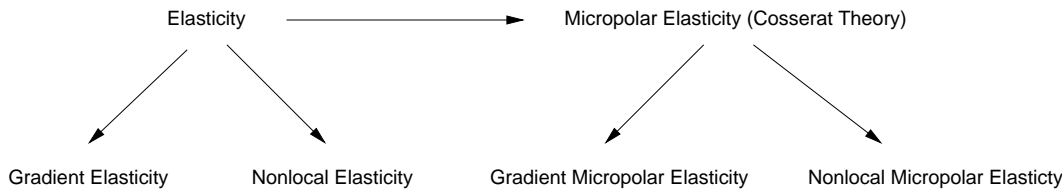,width=14cm}}
\end{picture}
}
\caption{Overview of generalizations of elasticity}
\label{fig:DOV}
\end{figure}

Another extension of the classical theory of elasticity is called strain gradient
elasticity and is very close to nonlocal elasticity.
The physical motivation to introduce gradient theories
was given by Kr\"oner~\cite{Kroener63,KD66} in the early sixties.
The strain gradient theories extend the classical elasticity with
additional strain gradient terms. Due to the gradients,
they must contain additional material constants with the dimension of
a length, and hyperstresses appear. The hyperstress tensor 
is a higher order stress tensor given in terms of
strain gradient terms. 
In particular, early theories of isotropic, higher-order gradient, linear elasticity 
were developed essentially by Mindlin~\cite{Mindlin64,Mindlin65,ME68},
Green and Rivlin~\cite{GR64a,GR64b}.
Such gradient theories contain strain gradient terms and no
rotation vector. In this way, they 
are different from theories with rotational degrees of freedom (Cosserat theory, 
micropolar elasticity). 
Only hyperstresses such as double or triple stresses
appear in strain gradient theories~\cite{Mindlin65}. 
Gradient elasticity was used to calculate the stress and strain fields
produced by dislocations and disclinations~\cite{GA99,Gutkin00,LM03a}. 
The gradient elasticity solutions have
no singularity in both the stress and the strain fields.  
On the other hand, in first gradient elasticity 
the double stresses still have singularities at the defect line~\cite{LM03a}.
We notice that in the special gradient elasticity used 
by Gutkin and Aifanis~\cite{GA96,GA97,GA99b} the stresses are still singular.
They only eliminated the strain singularities.  
Also, the gradient theories latter developed and 
used by Gutkin and Aifantis~\cite{GA99,Gutkin00,Aifan03} 
and the gradient theory by Lazar and Maugin~\cite{LM03a} 
are slightly different.
Gutkin and Aifantis used two different gradient coefficients and 
Lazar and Maugin used only one gradient coefficient.  

Eringen~\cite{Eringen84,Eringen02} introduced a theory of nonlocal micropolar elasticity (or
nonlocal Cosserat theory). In this theory the couple stress also has
a nonlocal constitutive relation in addition to the asymmetric 
force stress tensor. 
By means of nonlocal micropolar elasticity, the singularities in both the
force stresses and couple stresses can be eliminated~\cite{Povstenko98,PM97}. 
But, the micropolar distortion and wryness still possess singularities.
This was one motivation for Lazar and Maugin
to introduce a gradient micropolar elasticity~\cite{LM04a,LM04b}.
The gradient micropolar elasticity solutions eliminate these singularities
at the defect line.

Thus, in this paper we only 
consider various generalizations of classical linear elasticity
and not physically nonlinear elasticity. Nevertheless,
physically nonlinear elasticity is an alternative way to eliminate 
classical singularities at the dislocation line (see, e.g.,~\cite{zubov}).
But nonlinear elasticity is still scale-free and no internal lengths appear.

The aim of this paper is to give an overview of the solutions of 
dislocations in gradient elasticity, nonlocal elasticity, gradient
micropolar elasticity and nonlocal micropolar elasticity of Helmholtz type.
Gradient theory of Helmholtz type coincides with a special class of
first order gradient theory. Nonlocal elasticity of Helmholtz type means
that the nonlocal kernel is the Green function of the two-dimensional 
Helmholtz equation. We give the results for screw and edge dislocations. 
We consider only gradient theories and nonlocal theories with 
only one gradient coefficient and one nonlocal parameter, respectively.

\section{Generalized Elasticity of Helmholtz type}
\setcounter{equation}{0}

\subsection{Gradient elasticity of first order} 
In first gradient elasticity, the strain energy is assumed
to depend only on the elastic strain $E_{ij}$ and on the 
first gradients of it~\cite{Mindlin65,ME68,AA97,LM03a}
\begin{align}
\label{strain-en}
W=W(E_{ij},\pd_k E_{ij}),
\end{align}
where in the most general case of linear elasticity -- the incompatible elasticity --
the elastic strain 
is given by
\begin{align}
\label{strain-deco}
E_{ij}=\frac{1}{2}\,\big(\pd_j u_i+\pd_i u_j\big)-E^P_{ij}.
\end{align}
Here $u_i$ is the displacement vector and $E^P_{ij}$ denotes
the plastic strain.
The elastic strain (\ref{strain-deco}) is the symmetric part 
of the elastic distortion
\begin{align}
\label{dist-deco}
\beta_{ij}=\pd_j u_i-\beta^P_{ij},
\end{align}
where $\beta^P_{ij}$ is the plastic distortion.
As usual, the gradient of the displacement defines the total distortion
\begin{align}
\label{dist-T}
\beta^T_{ij}=\pd_j u_i.
\end{align}
With Eq.~(\ref{strain-deco}) the strain energy~(\ref{strain-en}) may be rewritten
in terms of gradients of the displacement and the plastic strain
according to
\begin{align}
\label{strain-en1b}
W=W(\pd_{(i}u_{j)}, \pd_k\pd_{(i}u_{j)},E^P_{ij},\pd_k E^P_{ij}).
\end{align}

In the theory of dislocations (see, e.g.,~\cite{Kroener81}), the dislocation tensor is defined by
\begin{align}
\label{DD}
\alpha_{ij}=\epsilon_{jkl}\pd_k\beta_{il}=-\epsilon_{jkl}\pd_k\beta^P_{il}
\end{align}
as incompatibility condition
and, on the other hand, the compatibility condition for the total distortion 
reads 
\begin{align}
\label{CC-1}
\epsilon_{jkl}\pd_k\beta^T_{il}=0.
\end{align}
By differentiating Eq.~(\ref{DD}) we obtain the 
translational Bianchi identity for $\alpha_{ij}$:
\begin{align}
\pd_j\alpha_{ij}=0.
\end{align}
If the plastic strain and plastic distortion are zero, 
we have compatible elastic strain and distortion which
are only given in terms of the gradient of the displacement
and, thus, the dislocation tensor (\ref{DD}) must be zero.
Such compatible strain gradient theories were investigated
by Mindlin~\cite{Mindlin64,Mindlin65,ME68}. 
Because we are interested in dislocations which cause plastic strain,
we must use an incompatible strain gradient theory~\cite{LM03a}. 
 
In gradient elasticity of Helmholtz type, 
the following expression of
the strain energy is postulated~\cite{AA97,LM03a}
\begin{align}
\label{en-sy}
W=\frac{1}{2}\,\sigma_{ij}E_{ij}+\frac{1}{2}\varepsilon^2\big(\pd_k\sigma_{ij}\big)\big(\pd_k E_{ij}\big),
\end{align}
where the parameter $\varepsilon$ is called the gradient coefficient 
which has the dimension of a length.
The pertinent stress tensors can be defined by taking the variation of $W$
\begin{align}
\label{CR1}
\sigma_{ij}&:=\frac{\pd W}{\pd E_{ij}},\\
\label{tau2}
\tau_{ijk}&:=\frac{\pd W}{\pd\big( \pd_k E_{ij}\big)}
        =\varepsilon^{2}\,\pd_k \sigma_{ij} 
\end{align}
with the Hooke law for an isotropic media 
\begin{align}
\label{HL1}
\sigma_{ij}=\lambda\,\delta_{ij} E_{kk}+ 2 \mu E_{ij},
\end{align}
where $\lambda$ and $\mu$ are the Lam{\'e} coefficients.
For nonnegative strain energy, $W\ge 0$, one has
\begin{align}
3\lambda+2\mu\ge 0,\quad \mu\ge 0,\quad \varepsilon^{2}\ge 0.
\end{align}
The stress tensor $\sigma_{ij}=\sigma_{ji}$ is a Cauchy-like stress tensor
and $\tau_{ijk}=\tau_{jik}$ is the double stress tensor.
Consequently, the double stress~(\ref{tau2}) is a simple gradient of the Cauchy stress.
Thus, it is a higher-order stress tensor.
Only $\varepsilon^{2}$ is a non-standard coefficient of the theory.

Further,
for vanishing external body forces, the force equilibrium equation 
can be obtained from the principle of 
virtual work as (variation with respect to the displacement)~\cite{Mindlin64}
\begin{align}
\label{EC2}
\pd_j\big(\sigma_{ij}-\pd_k\tau_{ijk}\big)=0
\end{align}
and with (\ref{tau2}) we find
\begin{align}
\label{}
\big(1-\varepsilon^{2}\Delta\big)\pd_j\sigma_{ij}=0,
\end{align}
where $\Delta$ denotes the Laplacian.
Finally, it is convenient to introduce
a quantity, which is called the total stress tensor, and is defined as 
\begin{align}
\label{CR1-t}
\tl\sigma_{ij}=\sigma_{ij}-\pd_k\tau_{ijk}      
              =\lambda\delta_{ij} E_{kk}+2\mu E_{ij}-\varepsilon^{2}\big(\lambda\delta_{ij}\Delta E_{kk}+2\mu \Delta E_{ij}\big).
\end{align}
With this definition
Eq.~(\ref{EC2}) takes the form
\begin{align}
\label{EC1}
\pd_j\tl\sigma_{ij}=0.
\end{align}
If we rewrite Eq.~(\ref{CR1-t}), we obtain for every component of the
Cauchy stress an inhomogeneous Helmholtz equation
\begin{align}
\label{stress-fe}
\big(1-\varepsilon^{2}\Delta\big)\sigma_{ij}=\tl\sigma {}_{ij}.
\end{align}
The RHS of (\ref{stress-fe}) is given in terms of the total stress tensor.

Using the inverse of Hooke's law for the stress $\sigma_{ij}$ and $\tl\sigma {}_{ij}$, 
it follows that the elastic strain can be determined from the equation
\begin{align}
\label{strain-fe}
\big(1-\varepsilon^{2}\Delta\big)E_{ij}=\tl E {}_{ij},
\end{align}
where $\tl E {}_{ij}$ is the classical strain 
tensor.

Using Eqs.~(\ref{strain-deco}) and (\ref{dist-deco}), we obtain the
coupled partial differential equation
\begin{align}
\label{PDE-coup}
\big(1-\varepsilon^{2}\Delta\big)\big[\pd_{(i}u_{j)}-\beta^P_{(ij)}\big]=
\pd_{(i}\tl u {}_{j)}-\tl \beta {}^P_{(ij)},
\end{align}
where $\tl u {}_i$ denotes the  
displacement field and $\tl \beta {}^P_{ij}$ is the 
plastic distortion in classical defect theory (see, e.g.,~\cite{deWit73b}).
Thus, if the following equations are satisfied
\begin{align}
\label{dist-HE}
&\big(1-\varepsilon^{2}\Delta\big)\beta_{ij}=\tl \beta {}_{ij},\\
\label{plast-HE}
&\big(1-\varepsilon^{2}\Delta\big)\beta^P_{ij}=\tl \beta {}^P_{ij},
\end{align} 
the equation for the displacement field,
\begin{align}
\label{u-HE}
\big(1-\varepsilon^{2}\Delta\big) u_{i}=\tl u {}_{i},
\end{align}
is valid for the incompatible case. 
Thus, for defects (dislocations, disclinations) 
the inhomogeneous parts of Eqs.~(\ref{plast-HE}) and (\ref{u-HE})
are fields with discontinuities.
The discontinuity of the displacement field of a defect 
is usually represented by a branch cut in order to make the multi-valued 
part to a single-valued one.
Thus this discontinuity is for mathematical convenience only. 
This can be accepted because the displacement and the plastic distortion
are not physical state variables. 
Therefore, the displacement and the plastic distortion depend on the 
choice of the branch cut.
For that reason, they are not unique.  

From Eqs.~(\ref{DD}), (\ref{dist-HE}) and (\ref{plast-HE}) 
we obtain 
\begin{align}
\label{DD-HE}
\big(1-\varepsilon^{2}\Delta\big)\alpha_{ij}=\tl \alpha {}_{ij},
\end{align} 
where $\tl \alpha_{ij}$ is the classical dislocation density tensor.
For a straight dislocation it has the form
\begin{align}
\tl \alpha_{ij}=b_i\otimes n_j\, \delta(x)\delta(y),
\end{align}
where $b_i$ and $n_j$ denote the Burgers vector and the direction of the dislocation line,
respectively. 
Then $\alpha_{ij}$ is the two-dimensional Green function of (\ref{DD-HE}) 
and reads
\begin{align}
\alpha_{ij}= \frac{1}{2\pi\varepsilon^2}\, b_i\otimes n_j\, K_0(r/\varepsilon)
\end{align}
where $r=\sqrt{x^2+y^2}$ and
$K_n$ is the modified Bessel function of the second kind and 
$n=0,1,\ldots$ denotes the order of this function.

The limit $\varepsilon\rightarrow 0$ is the limit from 
gradient elasticity theory to classical theory of elasticity.

\subsection{Nonlocal Elasticity}
The basic equations of linear, isotropic, nonlocal 
elastic solids, for the static case with vanishing body force,
are~\cite{Eringen77a,Eringen77b,Eringen83,Eringen02} 
\begin{align}
\label{FE-nl}
&\pd_j \sigma_{ij}=0,\\
\label{stress-nl}
&\sigma_{ij}(r)=\int_V \alpha(r-r')\,\sigma^{\text{(cl)}}_{ij}(r')\, \d v(r'),
\\
\label{HL2}
&\sigma^{\text{(cl)}}_{ij}=\lambda\,\delta_{ij} \epsilon^{\text{(cl)}}_{kk}
+2\mu\, \epsilon^{\text{(cl)}}_{ij}.
\end{align}
Here $\epsilon^{\text{(cl)}}_{ij}$ is the classical strain tensor,
$\sigma^{\text{(cl)}}_{ij}$, and $\sigma_{ij}$ are the classical and nonlocal stress tensors,
respectively.
In addition, $\lambda$ and $\mu$ are the Lam{\'e} constants
and $\alpha(r)$ is the `attenuation function' called  nonlocal kernel.
It is important to note that the local material coefficients
in Eqs.~(\ref{HL2}) 
are the same which appear in (\ref{HL1}).

Eringen and Ari~\cite{AE83} 
found for the two-dimensional case that an
excellent match of phonon dispersion was obtained, in the entire 
Brillouin zone, by
\begin{align}
\label{green}
\alpha(r-r')& =\frac{1}{2\pi\varepsilon^2}\,K_0\big( \sqrt{(x-x')^2+(y-y')^2}/\varepsilon\big),
\quad \varepsilon\ge 0.
\end{align}
$\varepsilon$ is called the parameter of nonlocality.
It is interesting to note that the nonlocal kernel~(\ref{green}) is
the Green function for the two-dimensional Helmholtz equation
\begin{align}
\big(1-\varepsilon^{2}\Delta\big)\alpha(r)=\delta(x)\delta(y).
\end{align}
Thus, we call (\ref{green}) nonlocal kernel of Helmholtz-type.
In this way, we deduce Eringen's so-called nonlocal constitutive relation
for a linear, isotropic solid with Green's function~(\ref{green}) 
as the nonlocal kernel. This kernel~(\ref{green}) has its maximum at $r=r'$ and
describes the nonlocal interaction. 
The normalization condition for the nonlocal kernel is given by
\begin{align}
\int_V\alpha(r-r')\,\d v(r')=1.
\end{align}
In the classical limit ($\varepsilon\rightarrow 0$), it becomes the two-dimensional 
Dirac delta function
\begin{align}
\lim_{\varepsilon\to 0}\alpha(r-r')= \delta(x-x')\delta(y-y').
\end{align}
In this limit, Eq.~(\ref{stress-nl}) gives the classical expression.
We notice that Eringen~\cite{Eringen83,Eringen02} found the two-dimensional kernel~(\ref{green}) 
by giving the best match with the Born-K{\'a}rm{\'a}n model of the 
atomic lattice dynamics and the atomistic dispersion curves.
He used the choice $e_0=0.39$ for the length,  
$\varepsilon=e_0\,a$,
where $a$ is an internal length (e.g. atomic lattice parameter)
and $e_0$ is a constant appropriate to each material. 

Applying the Helmholtz operator $(1-\varepsilon^2\Delta)$ to (\ref{stress-nl}),
we find the following inhomogeneous Helmholtz equation 
\begin{align}
\label{stress-fe-nl}
\big(1-\varepsilon^{2}\Delta\big) \sigma_{ij}=\sigma^{\text{(cl)}}_{ij},
\end{align}
where $\sigma^{\text{(cl)}}_{ij}$ is the stress tensor
obtained for the same traction boundary-value problem 
within the `classical' theory.
The factor $\varepsilon$ has the physical dimension of a length and 
it, therefore, defines an internal characteristic length.

It is obvious that a `nonlocal' strain tensor does not appear in
nonlocal elasticity. 
Thus, the strain and the displacement fields are the classical ones in
Eringen's theory of nonlocal elasticity. 
The same is true for the double stresses which occur 
in gradient elasticity. 
If we identify $\sigma^{\text{(cl)}}_{ij}=\tl\sigma_{ij}$,
Eqs.~(\ref{stress-fe}) and (\ref{stress-fe-nl}) coincide and the 
stress $\sigma_{ij}$ also fulfills $\pd_j\sigma_{ij}=0$ in gradient elasticity. 
Then the total stress tensor is equal to the classical stress tensor. 
Only, the stress tensor $\sigma_{ij}$ is the unknown field which must be found.

\subsection{Screw Dislocation}
We begin with the simplest case of a defect -- the straight screw dislocation.
First, we need the classical quantities because they act as 
inhomogeneous parts in gradient elasticity as well in nonlocal elasticity.
The classical stress is given by 
\begin{align}
\label{T-bg}
\tl\sigma_{z\varphi}=\pd_r\tl F=\frac{\mu b_z}{2\pi r}\, ,
\end{align}
with the Prandtl stress function
\begin{align}
\label{SF-scr}
\tl F=\frac{\mu b_z}{2\pi}\,\ln r\, .
\end{align}
The modified stress can be expressed in terms of a new stress function $F$
as follows
\begin{align}
\label{SFA-scr}
\sigma_{z\varphi}=\pd_r F\, .
\end{align}
For a screw dislocation
Eq.~(\ref{stress-fe}) takes the form
\begin{align}
\label{HE-F-scr}
\big(1-\varepsilon^{2}\Delta\big)F=\frac{\mu b_z}{2\pi}\, \ln r\, .
\end{align} 
The nonsingular solution of the modified stress function for a screw dislocation
is given by
\begin{align}
\label{SF-screw}
F=\frac{\mu b_z}{2\pi}\Big\{\ln r +K_0(r/\varepsilon)\Big\}\, ,
\end{align}
which is a superposition of the Prandtl stress function and a gradient part
depending on $\varepsilon$.
Thus, it is the two-dimensional Green function of the following partial differential equation
of fourth order (Helmholtz Laplace equation)
\begin{align}
\label{Green-screw}
\big(1-\varepsilon^{2}\Delta\big)\Delta F=\mu b_z \delta(x)\delta(y)\, .
\end{align}
Because of the fact that (\ref{SF-screw}) satisfies the inhomogeneous 
Helmholtz equation in which the Prandtl stress function is the inhomogeneous 
part, one might call (\ref{SF-screw}) 
the Helmholtz-modified Prandtl stress function.

\begin{figure}[t]\unitlength1cm
\centerline{
\begin{picture}(8,6)
\put(0.0,0.2){\epsfig{file=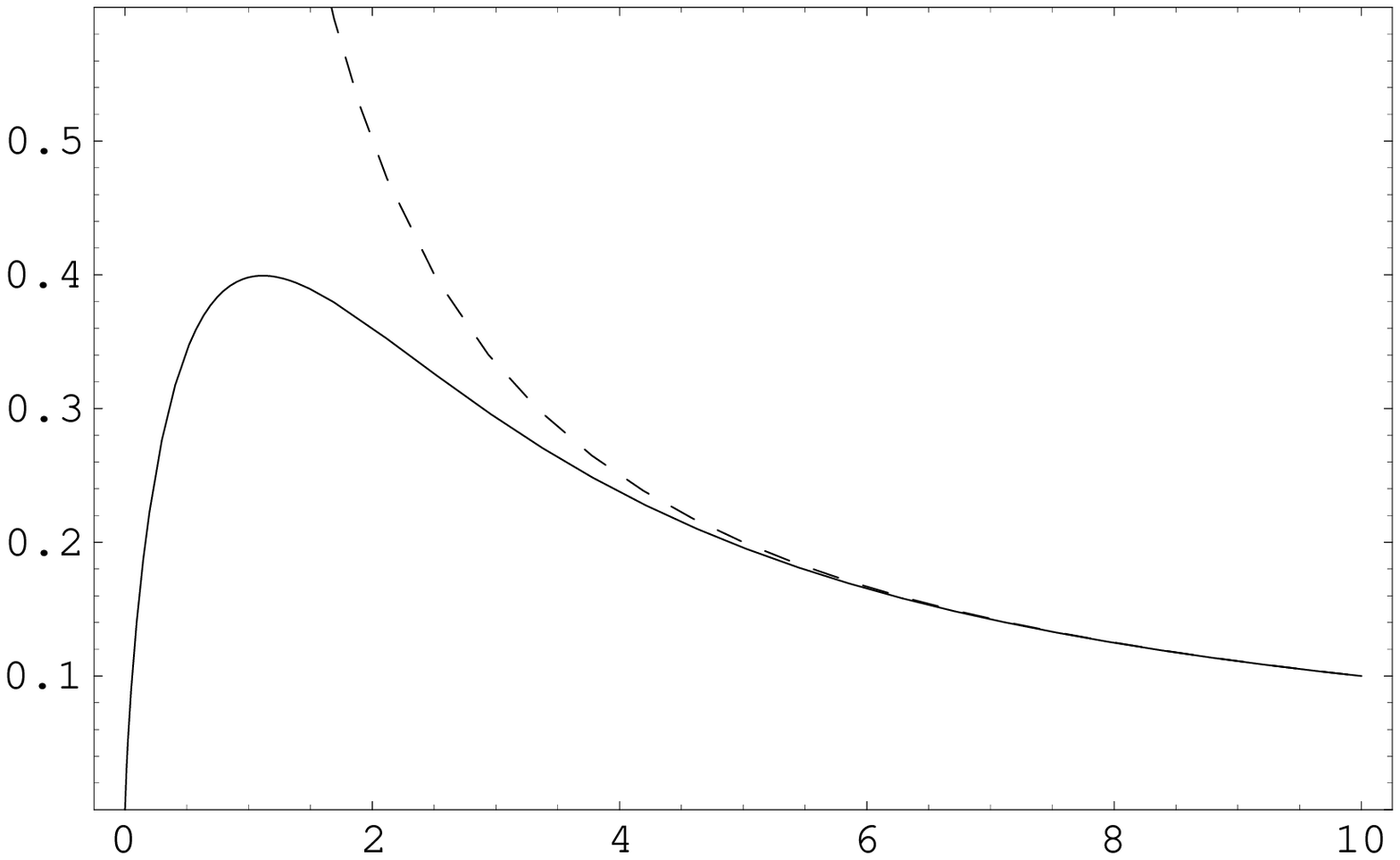,width=9cm}}
\put(4.5,0.0){$r/\varepsilon$}
\put(-1.0,4.5){$\sigma_{z\varphi}$}
\end{picture}
}
\caption{The stress $\sigma_{z\varphi}$  
is given in units of $\mu b_z/[2\pi\varepsilon]$.  
The dashed curve represents the classical result.}
\label{fig:T-scr}
\end{figure}

Using Eqs.~(\ref{SFA-scr}) and (\ref{SF-screw}), the stress field
is given by
\begin{align}
\label{T-cyl}
\sigma_{z\varphi}=\frac{\mu b_z}{2\pi}\,\frac{1}{r}
\Big\{1-\frac{r}{\varepsilon}\, K_1(r/\varepsilon)\Big\}\, .
\end{align}
Unlike the classical result, it does not possess any singularity at $r=0$.
It has a maximum $\sigma_{z\varphi}\simeq 0.399 \mu b_z/[2\pi\varepsilon]$ 
at about $r\simeq 1.114\, \varepsilon$. The plot of the stress 
versus $r/\varepsilon$ is shown in Fig.~\ref{fig:T-scr}. 
The stress~(\ref{T-cyl}) is also the solution
of Eq.~(\ref{stress-fe-nl}) in nonlocal elasticity. In nonlocal elasticity
this solution was found by Eringen~\cite{Eringen83}.
It is interesting to notice that 
a solution of a screw dislocation in nonlocal elasticity with
a nonlocal kernel which is the Green function of a diffusion-like equation
was earlier given by Eringen~\cite{Eringen77a}.

By means of the inverse of Hooke's law, the elastic strain is given by 
\begin{align}
\label{E-cyl}
E_{z\varphi}=\frac{b_z}{4\pi}\,\frac{1}{r}\Big\{1-\frac{r}{\varepsilon}\, K_1(r/\varepsilon)\Big\}\, ,
\end{align}
which is nonsingular at the dislocation line 
and has
its maximum $E_{z\varphi}\simeq 0.399 b_z/[2\pi\varepsilon]$ 
at $r\simeq 1.114\, \varepsilon$.
The components of stresses and strains in Cartesian coordinates 
are given in Refs.~\cite{GA96,GA99,Gutkin00,Lazar03,LM03a}. 

\subsection{Edge Dislocation}
The components of the classical stress of a straight edge dislocation with
Burgers vector along the $x$-direction read in cylindrical coordinates
\begin{align}
\label{SFA-edge-cl}
\tl\sigma_{rr}=\frac{1}{r}\,\pd_r \tl f+\frac{1}{r^2}\,\pd^2_{\varphi\varphi} \tl f,\quad
\tl\sigma_{\varphi\varphi}=\pd^2_{rr} \tl f,\quad
\tl\sigma_{r\varphi}=-\pd_r\left(\frac{1}{r}\,\pd_\varphi\tl f\right),\quad
\tl\sigma_{zz}=\nu(\tl\sigma_{rr}+\tl\sigma_{\varphi\varphi}),
\end{align}
where the Airy stress function is given by
\begin{align}
\label{Airy1}
\tl f=-\frac{\mu b_x}{2\pi(1-\nu)}\,
\sin\varphi\, \big(r \ln r\big)\, .
\end{align}
For the modified stress of an edge dislocation, the stresses are given in 
terms of a stress function $f$ similar to~(\ref{SFA-edge-cl})
\begin{align}
\label{SFA-edge}
\sigma_{rr}=\frac{1}{r}\,\pd_r f+\frac{1}{r^2}\,\pd^2_{\varphi\varphi} f,\quad
\sigma_{\varphi\varphi}=\pd^2_{rr} f,\quad
\sigma_{r\varphi}=-\pd_r\left(\frac{1}{r}\,\pd_\varphi f\right),\quad
\sigma_{zz}=\nu(\sigma_{rr}+\sigma_{\varphi\varphi}).
\end{align}
For an edge dislocation~(\ref{stress-fe}) has the following form
\begin{align}
\label{f_fe}
\big(1-\varepsilon^{2}\Delta\big)f=-\frac{\mu b_x}{2\pi(1-\nu)}\,
\sin\varphi\,\big(r\ln r\big) ,
\end{align}
with the Airy function as inhomogeneous part.
The nonsingular solution of (\ref{f_fe}) is given by~\cite{Lazar03a}
\begin{align}
\label{SF-edge}
f=-\frac{\mu b_x}{2\pi(1-\nu)}\, \sin\varphi \Big\{r\ln r 
+\frac{2\varepsilon^2}{r}\Big(1-\frac{r}{\varepsilon}\, K_1(r/\varepsilon)\Big)\Big\}. 
\end{align}
It is a superposition of the Airy stress function and a gradient part.
We may call the stress function~(\ref{SF-edge}) 
a Helmholtz-modified Airy stress function since
it fulfills (Helmholtz bi-Laplace equation)
\begin{align}
\label{Green-edge}
\big(1-\varepsilon^{2}\Delta\big)\Delta\Delta f=
-\frac{2\mu b_x}{(1-\nu)}\,\pd_y \delta(x)\delta(y) .
\end{align}
Thus, it is a Green function of this partial differential equation of sixth
order.

\begin{figure}[p]\unitlength1cm
\vspace*{-1.0cm}
\centerline{
(a)
\begin{picture}(8,6)
\put(0.0,0.2){\epsfig{file=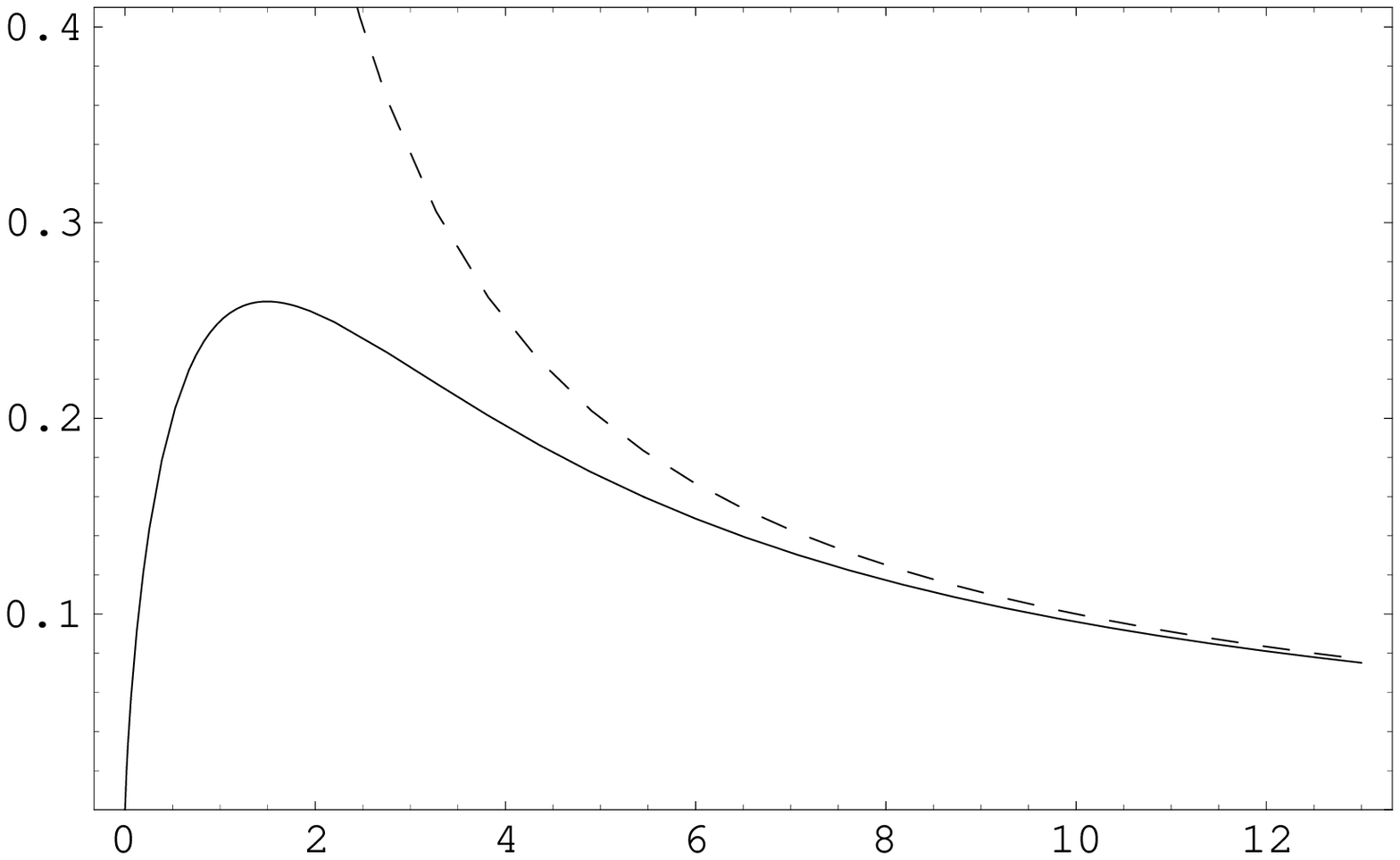,width=9cm}}
\put(4.5,0.0){$r/\varepsilon$}
\put(-1.0,4.5){$\sigma_{rr}$}
\end{picture}
}
\centerline{
(b)
\begin{picture}(8,6)
\put(0.0,0.2){\epsfig{file=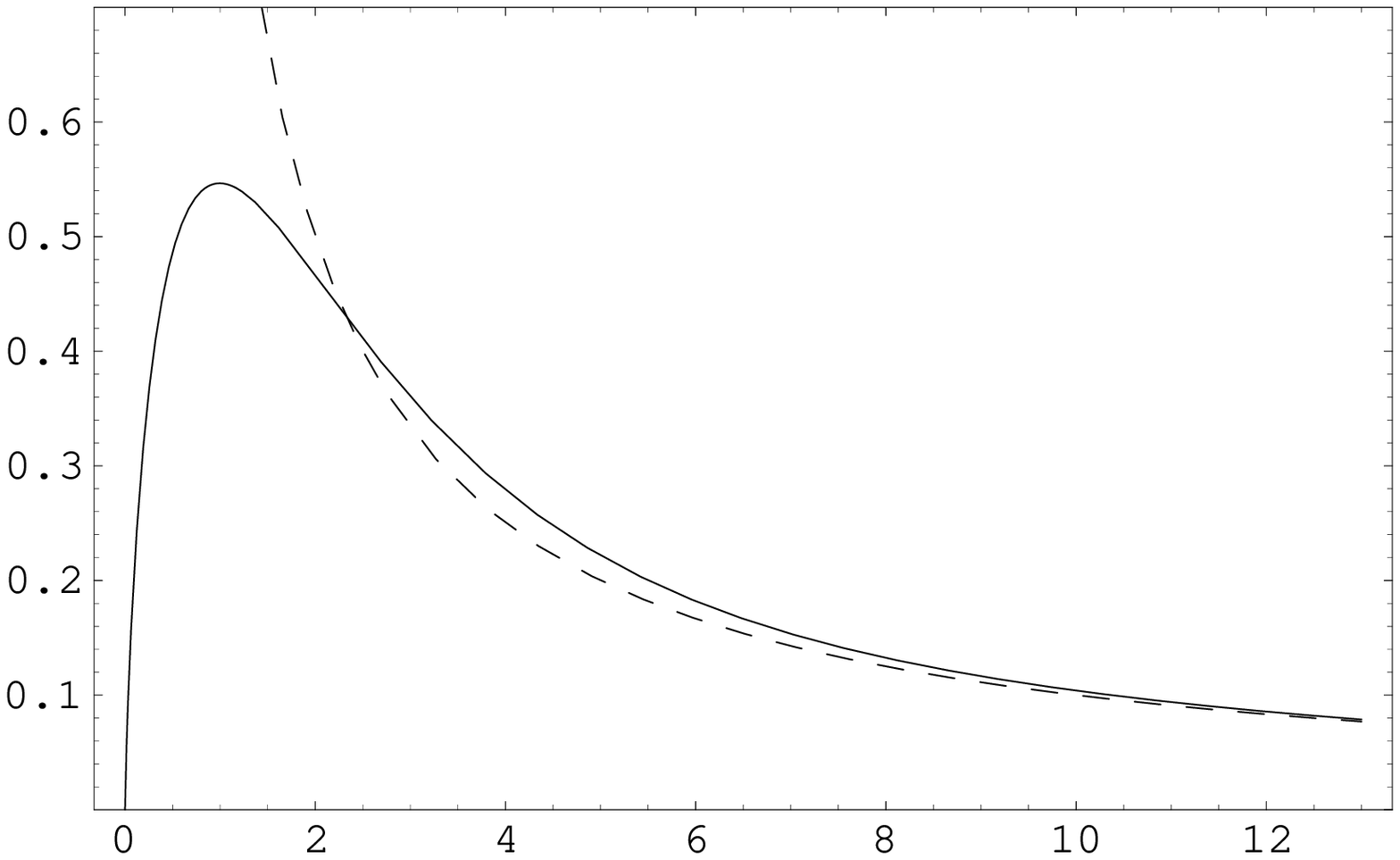,width=9cm}}
\put(4.5,0.0){$r/\varepsilon$}
\put(-1.0,4.5){$\sigma_{\varphi\varphi}$}
\end{picture}
}
\centerline{
(c)
\begin{picture}(8,6)
\put(0.0,0.2){\epsfig{file=Tscr.eps,width=9cm}}
\put(4.5,0.0){$r/\varepsilon$}
\put(-1.0,4.4){$\sigma_{zz}$}
\end{picture}
}
\caption{The components of stress:
(a)~$\sigma_{rr}$ and (b)~$\sigma_{\varphi\varphi}$  
are given in units of $\mu b_z/[2\pi(1-\nu)\varepsilon]$
and  (c)~$\sigma_{zz}$ is given in units of $\mu b_z\nu /[\pi(1-\nu)\varepsilon]$
for $\varphi=3\pi/2$.  
The dashed curves represent the classical components.}
\label{fig:T-edge}
\end{figure}

By means of Eqs.~(\ref{SFA-edge}) and (\ref{SF-edge}), the stress field
of an edge dislocation is calculated as
\begin{align}
\label{T_rr}
\sigma_{rr}&=-\frac{\mu b_x}{2\pi(1-\nu)}\, 
\frac{\sin\varphi}{r}\Big\{1-\frac{4\varepsilon^2}{r^2}+2 K_2(r/\varepsilon)\Big\},\\
\label{T_rp}
\sigma_{r\varphi}&=\frac{\mu b_x}{2\pi(1-\nu)}\, 
\frac{\cos\varphi}{r}\Big\{1-\frac{4\varepsilon^2}{r^2}+2 K_2(r/\varepsilon)\Big\},\\
\label{T_pp}
\sigma_{\varphi\varphi}&=-\frac{\mu b_x}{2\pi(1-\nu)}\, 
\frac{\sin\varphi}{r}\Big\{1+\frac{4\varepsilon^2}{r^2}
-2 K_2(r/\varepsilon)-2\frac{r}{\varepsilon}\, K_1(r/\varepsilon)\Big\},\\
\label{T_zz}
\sigma_{zz}&=-\frac{\mu b_x\nu }{\pi(1-\nu)}\, 
\frac{\sin\varphi}{r}\Big\{1-\frac{r}{\varepsilon}\, K_1(r/\varepsilon)\Big\},
\end{align}
and the trace of the stress tensor reads
\begin{align}
\label{T_kk}
\sigma_{kk}=-\frac{\mu b_x(1+\nu)}{\pi(1-\nu)}\, 
\frac{\sin\varphi}{r}\Big\{1-\frac{r}{\varepsilon}\, K_1(r/\varepsilon)\Big\}.
\end{align}
The stress is plotted over $r/\varepsilon$ in Fig.~\ref{fig:T-edge}.
Based on the present solution~(\ref{T_rr})--(\ref{T_kk}) we observe 
that the stress fields are not singular at $r=0$. In fact, they are zero at $r=0$. 
The radial extremum values are:
$\sigma_{rr}\simeq -0.259  \mu b_x/[2\pi(1-\nu)\varepsilon]\sin\varphi$ at $r\simeq 1.494\,\varepsilon$, 
$\sigma_{r\varphi}\simeq 0.259 \mu b_x/[2\pi(1-\nu)\varepsilon]\cos\varphi$ at $r\simeq 1.494\, \varepsilon$, 
$\sigma_{\varphi\varphi}\simeq -0.546 \mu b_x/[2\pi(1-\nu)\varepsilon]\sin\varphi$ at $r\simeq 0.546\,\varepsilon$, 
$\sigma_{zz}\simeq -0.399 \mu b_x\nu /[\pi(1-\nu)\varepsilon]\sin\varphi$ at $r\simeq 1.114\,\varepsilon$ and 
$\sigma_{kk}\simeq -0.399 \mu b_x(1+\nu) /[\pi(1-\nu)\varepsilon]\sin\varphi$ at $r\simeq 1.114\,\varepsilon$.
We note that the stresses~(\ref{T_rr})--(\ref{T_kk}) are also 
solution of Eq.~(\ref{stress-fe-nl}) in the nonlocal elasticity.
The solution of an edge dislocation given by Eringen~\cite{Eringen77b}
does not correspond to the nonlocal kernel~(\ref{green}).
It is the solution in nonlocal elasticity with the nonlocal kernel
which is Green's function of a diffusion-like equation. Therefore,
it is different.

The elastic strain of a straight edge dislocation 
is given by
\begin{align}
\label{E_rr}
E_{rr}&=-\frac{b_x}{4\pi(1-\nu)}\, 
\frac{\sin\varphi}{r}\Big\{(1-2\nu)-\frac{4\varepsilon^2}{r^2}+2 K_2(r/\varepsilon)
+2\nu\frac{r}{\varepsilon}\, K_1(r/\varepsilon)\Big\},\\
\label{E_rp}
E_{r\varphi}&=\frac{b_x}{4\pi(1-\nu)}\, 
\frac{\cos\varphi}{r}\Big\{1-\frac{4\varepsilon^2}{r^2}+2 K_2(r/\varepsilon)\Big\},\\
\label{E_pp}
E_{\varphi\varphi}&=-\frac{b_x}{4\pi(1-\nu)}\, 
\frac{\sin\varphi}{r}\Big\{(1-2\nu)+\frac{4\varepsilon^2}{r^2}
-2 K_2(r/\varepsilon)-2(1-\nu)\frac{r}{\varepsilon}\, K_1(r/\varepsilon)\Big\}.
\end{align}
The dilatation reads
\begin{align}
\label{E_kk}
E_{kk}=-\frac{b_x(1-2\nu)}{2\pi(1-\nu)}\, 
\frac{\sin\varphi}{r}\Big\{1-\frac{r}{\varepsilon}\, K_1(r/\varepsilon)\Big\}.
\end{align}
\begin{figure}[t]\unitlength1cm
\centerline{
(a)
\begin{picture}(8,6)
\put(0.0,0.2){\epsfig{file=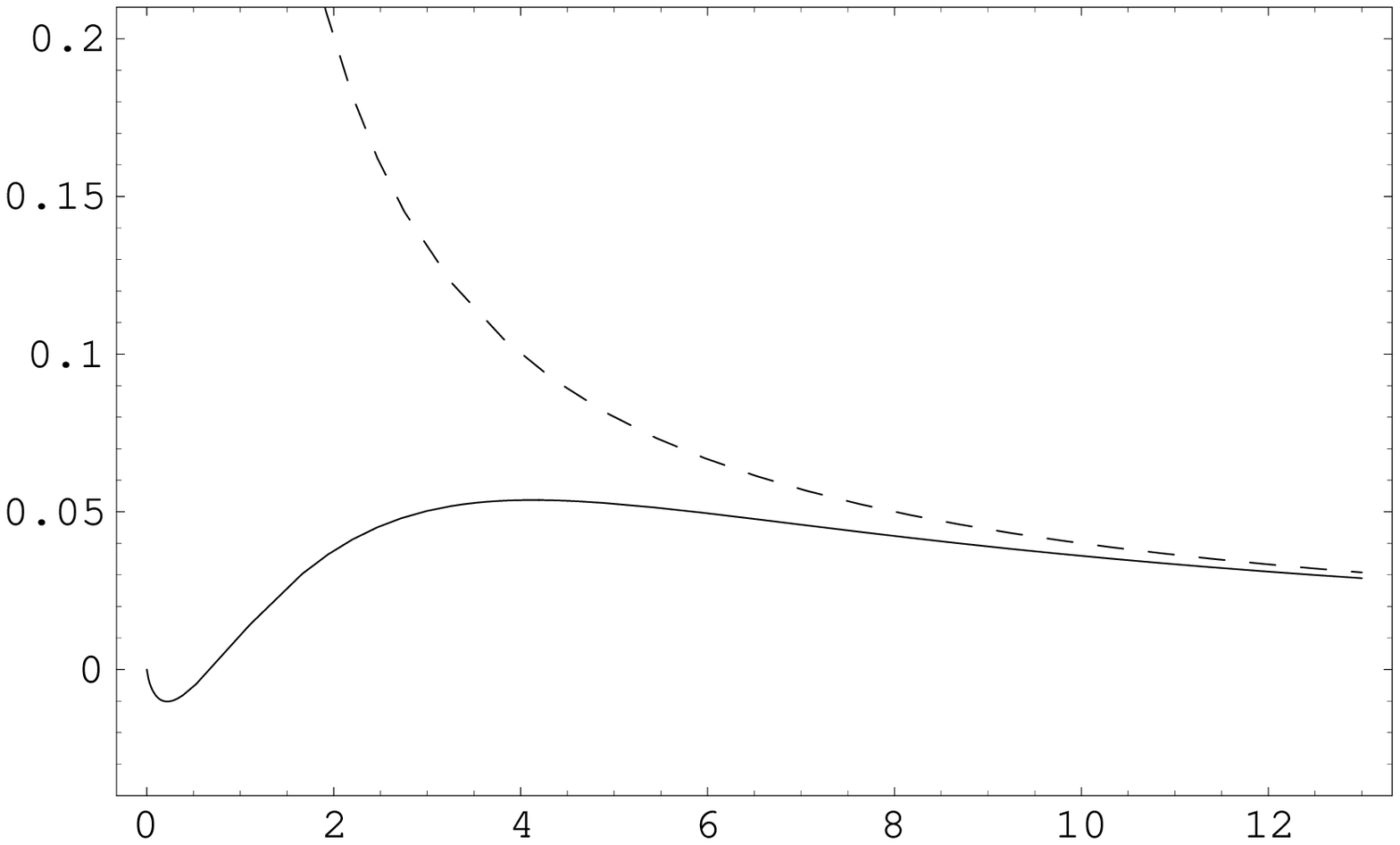,width=9cm}}
\put(4.5,0.0){$r/\varepsilon$}
\put(-1.0,4.5){$E_{rr}$}
\end{picture}
}
\centerline{
(b)
\begin{picture}(8,6)
\put(0.0,0.2){\epsfig{file=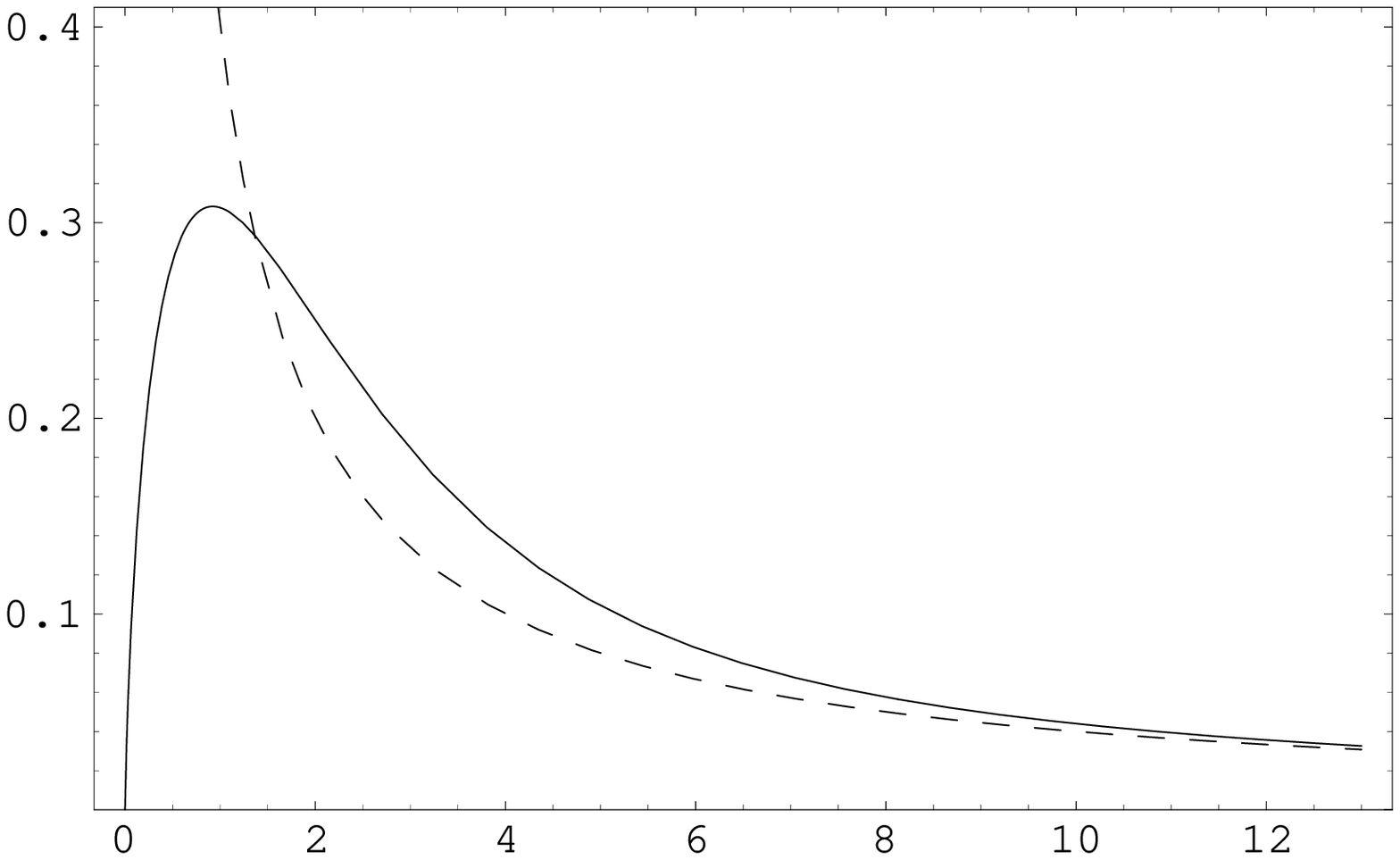,width=9cm}}
\put(4.5,0.0){$r/\varepsilon$}
\put(-1.0,4.5){$E_{\varphi\varphi}$}
\end{picture}
}
\caption{The components of strain:
(a)~$E_{rr}$ and (b)~$E_{\varphi\varphi}$  
are given in units of $ b_z/[4\pi(1-\nu)\varepsilon]$
for $\nu=0.3$ and $\varphi=3\pi/2$.  
The dashed curves represent the classical components.}
\label{fig:E-edge}
\end{figure}
Again the strain fields~(\ref{E_rr})--(\ref{E_kk}) are not singular at $r=0$. 
They are zero at the dislocation line. 
They have the following radial extremum values ($\nu=0.3$):
The strain $E_{rr}$ has two extremum values of opposite signs (see~Fig.~\ref{fig:E-edge}a), namely,
$E_{rr}\simeq 0.010 b_x/[4\pi(1-\nu)\varepsilon]\sin\varphi$ at $r\simeq 0.218\,\varepsilon$
and 
$E_{rr}\simeq -0.054 b_x/[4\pi(1-\nu)\varepsilon]\sin\varphi$ at $r\simeq 4.130\,\varepsilon$.
In addition, it is equal to zero at $r\simeq 0.677\,\varepsilon$. 
On the other hand, 
$E_{r\varphi}\simeq 0.259 b_x/[4\pi(1-\nu)\varepsilon]\cos\varphi$ at $r\simeq 1.494\, \varepsilon$, 
$E_{\varphi\varphi}\simeq -0.308 b_x/[4\pi(1-\nu)\varepsilon]\sin\varphi$ at $r\simeq 0.922\, \varepsilon$ 
(see Fig.~\ref{fig:E-edge}b)
and
$E_{kk}\simeq -0.399 \mu b_x(1-2\nu) /[2\pi(1-\nu)\varepsilon]\sin\varphi$ at $r\simeq 1.114\,\varepsilon$.
Unlike the classical solution, $E_{rr}$ is significantly smaller than $E_{\varphi\varphi}$
in the dislocation core region.
The components of the stress and strain tensors in Cartesian coordinates 
can be found in Refs.~\cite{GA97,GA99,Gutkin00,Lazar03a,Lazar03,LM03a}.

\section{Generalized Micropolar Elasticity of Helmholtz type}
\setcounter{equation}{0}

\subsection{Gradient Micropolar Elasticity of First Order}
In first gradient micropolar elasticity (or gradient Cosserat theory), 
the strain energy is assumed
to depend only on the micropolar distortion $\gamma_{ij}$,
the micropolar bend-twist $\kappa_{ij}$ and on the 
first gradients of them~\cite{LM04a,LM04b}
\begin{align}
\label{strain-en2}
W=W(\gamma_{ij},\kappa_{ij},\pd_k \gamma_{ij},\pd_k \kappa_{ij}),
\end{align}
where in the most general case -- the incompatible micropolar elasticity --
the elastic micropolar distortion is given by~\cite{Eringen99,LM04a}
\begin{align}
\label{gam-deco}
\gamma_{ij}=\pd_j u_i+\epsilon_{ijk}\varphi_k -\gamma^P_{ij}.
\end{align}
Here $\varphi_i$ is the micro-rotation vector 
and $\gamma^P_{ij}$ denotes the plastic 
micropolar distortion. In addition,
the elastic micropolar bend-twist (or wryness) reads~\cite{Eringen99,LM04a}
\begin{align} 
\label{kap-deco}
\kappa_{ij}=\pd_j \varphi_i- \kappa^P_{ij},
\end{align}
where $\kappa^P_{ij}$ denotes the plastic micropolar bend-twist tensor.
The total deformation quantities are given by
\begin{align}
\label{gam-T}
\gamma^T_{ij}&=\pd_j u_i+\epsilon_{ijk}\varphi_k ,\\
\label{kap-T}
\kappa^T_{ij}&=\pd_j \varphi_i.
\end{align}
For dislocation and disclination it is necessary to use the incompatible
gradient theory because both defects cause plastic fields.
Using Eqs.~(\ref{gam-deco}) and (\ref{kap-deco}), the 
strain energy~(\ref{strain-en2}) might be written according to
\begin{align}
\label{strain-en2b}
W=W(\pd_j u_i, \pd_k\pd_j u_i, \varphi_i,\pd_j\varphi_i,\pd_k\pd_j\varphi_i,
\gamma^P_{ij},\pd_k\gamma^P_{ij},\kappa^P_{ij},\pd_k \kappa^P_{ij}).
\end{align}
Thus, the expression~(\ref{strain-en2b}) contains gradients of the displacement
and rotation vectors in addition to gradients of the plastic fields.

By means of the plastic fields and the incompatible elastic ones 
the dislocation and disclination densities can be calculated as~\cite{Eringen99,LM04a}.
\begin{align}
\label{disl-d}
&\alpha_{ij}=\epsilon_{jkl}\big(\pd_k\gamma_{il}-\epsilon_{ilm}\kappa_{mk}\big)              
=-\epsilon_{jkl}\big(\pd_k\gamma^P_{il}-\epsilon_{ilm}\kappa^P_{mk}\big),\\
\label{discl-d}
&\Theta_{ij}=\epsilon_{jkl}\pd_k\kappa_{il}=-\epsilon_{jkl}\pd_k\kappa^P_{il}.
\end{align}
By differentiating Eq.~(\ref{disl-d}) and Eq.~(\ref{discl-d}) we 
obtain the translational and rotational Bianchi identities for the dislocation
and disclination density tensors in the theory of micropolar elasticity:
\begin{align}
\pd_j\alpha_{ij}-\epsilon_{ijk}\Theta_{jk}&=0,\\
\pd_j\Theta_{ij}&=0.
\end{align}
The compatibility conditions are
\begin{align}
\label{CC2}
\epsilon_{jkl}\big(\pd_k\gamma^T_{il}-\epsilon_{ilm}\kappa^T_{mk}\big)&=0,\\
\label{CC3}
\epsilon_{jkl}\pd_k\kappa^T_{il}&=0.
\end{align}

When disclinations are absent, the 
disclination density tensor $\Theta_{ij}$ and the plastic 
micropolar bend-twist tensor $\kappa^P_{ij}$ must be zero and
we have $\kappa_{ij}=\pd_j\varphi_i$ and $\gamma^P_{ij}=\beta^P_{ij}$.
Then Eq.~(\ref{discl-d}) is the compatibility condition for $\kappa_{ij}$. 
The elastic micropolar distortion~(\ref{gam-deco}) may be rewritten in the
form
\begin{align}
\gamma_{ij}=\beta_{ij}+\epsilon_{ijk}\varphi_k.
\end{align}
Then the dislocation density tensor simplifies to
\begin{align} 
\alpha_{ij}=\epsilon_{jkl}\pd_k\beta_{il}=-\epsilon_{jkl}\pd_k\beta^P_{il} .
\end{align}

In gradient micropolar elasticity of Helmholtz type the strain energy, $W$, 
is assumed to be given as~\cite{LM04a,LM04b} 
\begin{align}
\label{strain-en3}
W=\frac{1}{2}\,\sigma_{ij}\gamma_{ij}+\frac{1}{2}\,\mu_{ij}\kappa_{ij}
   +\frac{1}{2}\varepsilon^2\big(\pd_k\sigma_{ij}\big)\big(\pd_k\gamma_{ij}\big)
   +\frac{1}{2}\varepsilon^2\big(\pd_k\mu_{ij}\big)\big(\pd_k\kappa_{ij}\big),
\end{align}
where $\kappa$ is the only one gradient coefficient.
By means of the strain energy~(\ref{strain-en2}), the asymmetric stresses are defined by
\begin{align}
\sigma_{ij}&:=\frac{\pd W}{\pd\gamma_{ij}},\\
\mu_{ij}&:=\frac{\pd W}{\pd\kappa_{ij}},\\
\label{DFS}
\tau_{ijk}&:=\frac{\pd W}{\pd\big( \pd_k\gamma_{ij}\big)}
            =\varepsilon^2\, \pd_k \sigma_{ij},\\
\label{DCS}
\lambda_{ijk}&:=\frac{\pd W}{\pd\big( \pd_k\kappa_{ij}\big)}
                =\varepsilon^2\, \pd_k \mu_{ij}.
\end{align}
Here $\sigma_{ij}$ and $\mu_{ij}$ are the force stress tensor and couple stress
tensor, respectively. The higher order stresses $\tau_{ijk}$ and $\lambda_{ijk}$ are called
double force stress and double couple stress, respectively.

The constitutive relations for 
isotropic micropolar elasticity (see, e.g.,~\cite{Eringen02,Nowacki74,Nowacki86}) 
have the form
\begin{align}
\label{CE1}
&\sigma_{ij}=\lambda\,\delta_{ij}\gamma_{kk}+(\mu+\eta)\gamma_{ij}+(\mu-\eta)\gamma_{ji},\\
\label{CE2}
&\mu_{ij}=\alpha\,\delta_{ij}\kappa_{kk}+(\beta+\gamma)\kappa_{ij}+(\beta-\gamma)\kappa_{ji}.
\end{align}
Here $\lambda$, $\mu$, $\eta$, $\alpha$, $\beta$ and $\gamma$ are the 6 material
constants of the isotropic Cosserat continuum.
For nonnegative strain energy, $W\ge 0$, we have
\begin{align}
&3\lambda+2\mu\ge 0,\quad \mu\ge 0,\quad\eta\ge 0,\nonumber\\
&3\alpha+2\beta\ge 0,\quad \beta\ge 0,\quad\gamma\ge 0
\end{align}
and 
\begin{align}
\varepsilon^{2}\ge 0.
\end{align}
Using the six material constants of a Cosserat continuum (micropolar medium), 
two characteristic 
lengths $l$ and $h$ can be defined by~\cite{Nowacki74,Nowacki86}
\begin{align}
l^2=\frac{(\mu+\eta)(\beta+\gamma)}{4\mu\,\eta},\qquad
h^2=\frac{\alpha+2\beta}{4\eta},
\end{align}
which are Cosserat intrinsic lengths.
The above equations~(\ref{CE1}) and (\ref{CE2}) may be expressed 
in terms of stresses
\begin{align}
\label{CE1-inv}
&\gamma_{ij}=\lambda'\,\delta_{ij}\sigma_{kk}+(\mu'+\eta')\sigma_{ij}+(\mu'-\eta')\sigma_{ji},\\
\label{CE2-inv}
&\kappa_{ij}=\alpha'\,\delta_{ij}\mu_{kk}+(\beta'+\gamma')\mu_{ij}+(\beta'-\gamma')\mu_{ji}.
\end{align}
Here
\begin{align}
&2\mu'=\frac{1}{2\mu},\qquad
2\eta'=\frac{1}{2\eta},\qquad
2\beta'=\frac{1}{2\beta},\qquad
2\gamma'=\frac{1}{2\gamma},\\
&\lambda'=-\frac{\lambda}{2\mu(3\lambda+2\mu)}
        ,\qquad
\alpha'=-\frac{\alpha}{2\beta(3\alpha+2\beta)}.
\end{align}

After variation of~(\ref{strain-en2}) with respect to the displacement $u_i$ and the
rotation vector $\varphi_i$, the following equilibrium conditions
follow
\begin{align}
\label{f-eq}
&\pd_j\big(\sigma_{ij}-\pd_k\tau_{ijk}\big)=0,\\
\label{m-eq}
&\pd_j\big(\mu_{ij}-\pd_k\lambda_{ijk}\big)-\epsilon_{ijk}\big(\sigma_{jk}-\pd_l\tau_{jkl}\big)=0.
\end{align}
We may define the  
total stress and total couple stress tensors 
\begin{align}
\label{stress-t}
&\tl\sigma_{ij}=\sigma_{ij}-\pd_k\tau_{ijk},\\
\label{cs-t}
&\tl\mu_{ij}=\mu_{ij}-\pd_k\lambda_{ijk}.
\end{align}
So, Eqs.~(\ref{f-eq}) and ({\ref{m-eq}) take the form 
\begin{align}
\label{1NI-cl}
&\pd_j\tl\sigma_{ij}=0,\\
\label{2NI-cl}
&\pd_j \tl\mu_{ij}-\epsilon_{ijk}\tl\sigma_{jk}=0
\end{align}
or in terms of $\sigma_{ij}$ and $\mu_{ij}$
\begin{align}
\label{1NI-gr}
&\big(1-\varepsilon^{2}\Delta\big)\pd_j\sigma_{ij}=0,\\
\label{2NI-gr}
&\big(1-\varepsilon^{2}\Delta\big)\big(\pd_j \mu_{ij}-\epsilon_{ijk}\sigma_{jk}\big)=0.
\end{align}
From Eqs.~(\ref{stress-t}), (\ref{cs-t}), (\ref{DFS}) and (\ref{DCS})  
the force stress tensor $\sigma_{ij}$ and couple stress tensor $\mu_{ij}$ 
satisfy the inhomogeneous Helmholtz equations
\begin{align}
\label{fstress-HE}
&\big(1-\varepsilon^{2}\Delta\big)\sigma_{ij}=\tl\sigma {}_{ij},\\
\label{mstress-HE}
&\big(1-\varepsilon^{2}\Delta\big)\mu_{ij}=\tl\mu {}_{ij},
\end{align}
where 
the RHS are given in terms of 
total stress and couple stress tensors.

Using Eqs. (\ref{CE1-inv}) and (\ref{CE2-inv}), the inverse 
of Hooke's law for the stress $\sigma_{ij}$ and $\tl\sigma {}_{ij}$, 
it follows that the elastic strain can be determined from the equation
\begin{align}
\label{gam-HE}
&\big(1-\varepsilon^{2}\Delta\big)\gamma_{ij}=\tl \gamma {}_{ij},\\
\label{kap-HE}
&\big(1-\varepsilon^{2}\Delta\big) \kappa_{ij}=\tl \kappa {}_{ij},
\end{align}
where $\tl \gamma {}_{ij}$ and $\tl \kappa {}_{ij}$ are fields
in `classical' micropolar elasticity.
These are 
coupled partial differential equations. 
In fact, substituting Eqs.~(\ref{gam-deco}) and (\ref{kap-deco}) in
(\ref{gam-HE}) and (\ref{kap-HE}), we obtain 
\begin{align}
\label{PDE-coup1}
&\big(1-\varepsilon^{2}\Delta\big)
\big[\pd_{i}u_{j}+\epsilon_{ijk}\varphi_k-\gamma^P_{ij}\big]=
\pd_{i}\tl u_{j}+\epsilon_{ijk}\tl \varphi_k-\tl\gamma {}^P_{ij},\\
\label{PDE-coup2}
&\big(1-\varepsilon^{2}\Delta\big)
\big[\pd_{i}\varphi_{j}-\kappa^P_{ij}\big]=
\pd_{i}\tl \varphi_{j}-\tl\kappa {}^P_{ij},
\end{align}
where $\tl \varphi {}_i$ denotes the  
rotation field, $\tl \gamma {}^P_{ij}$ is the 
plastic micropolar distortion in defect theory  
(see, e.g.,~\cite{Eringen99}).
Thus, if and if only the following equations are fulfilled
\begin{align}
\label{plast-HE1}
&\big(1-\varepsilon^{2}\Delta\big)\gamma^P_{ij}=\tl \gamma {}^P_{ij},\\
\label{plast-HE2}
&\big(1-\varepsilon^{2}\Delta\big)\kappa^P_{ij}=\tl \kappa {}^P_{ij},
\end{align} 
the equations for the displacement and rotation fields,
\begin{align}
\label{u-HE2}
&\big(1-\varepsilon^{2}\Delta\big) u_{i}=\tl u {}_{i},\\
\label{rot-HE}
&\big(1-\varepsilon^{2}\Delta\big) \varphi_{i}=\tl \varphi {}_{i},
\end{align}
are valid for the incompatible case. 
Since $\tl \kappa {}^P_{ij}$ and $\kappa^P_{ij}$ are 
the plastic fields of disclinations,  
they are zero for dislocations.
In this way, Eq.~(\ref{rot-HE} follows directly from (\ref{PDE-coup2})
for dislocations.
Thus, in dislocation theory only
the inhomogeneous parts of Eqs.~(\ref{u-HE2}) and (\ref{plast-HE1})
are fields with discontinuities.

In addition, we obtain for defects in first gradient micropolar elasticity 
from Eqs.~(\ref{disl-d}), (\ref{discl-d}), (\ref{gam-HE}) and (\ref{kap-HE}) 
\begin{align}
\label{DD-HE2}
&\big(1-\varepsilon^{2}\Delta\big)\alpha_{ij}=\tl \alpha {}_{ij},\\
&\big(1-\varepsilon^{2}\Delta\big)\Theta_{ij}=\tl \Theta {}_{ij},
\end{align} 
where $\tl \Theta_{ij}$ is the classical disclination density tensor.
For a straight disclination it reads
\begin{align}
\tl \Theta_{ij}=\Omega_i\otimes n_j\, \delta(x)\delta(y),
\end{align}
where $\Omega_i$ denotes the Frank vector. 
Consequently, the dislocation density and the disclination density tensors of
a straight defect have in first gradient micropolar elasticity the following
form 
\begin{align}
\alpha_{ij}&= \frac{1}{2\pi\varepsilon^2}\, b_i\otimes n_j\, K_0(r/\varepsilon)\\
\Theta_{ij}&=\frac{1}{2\pi\varepsilon^2}\, \Omega_i\otimes n_j\, K_0(r/\varepsilon).
\end{align}

In the limit $\varepsilon\rightarrow 0$, we recover micropolar elasticity.
Thus, it is the `classical' limit in gradient micropolar elasticity.

\subsection{Nonlocal Micropolar Elasticity}
The basic equations in the isotropic theory of nonlocal micropolar elasticity 
are given  by~\cite{Eringen84,Eringen02}
\begin{align}
\label{FE-NL}
&\pd_j\sigma_{ij}=0,\\
\label{ME-NL}
&\pd_j \mu_{ij}-\epsilon_{ijk}\sigma_{jk}=0,\\
\label{fstress-nl}
&\sigma_{ij}(r)=\int_V \alpha(r-r')\,\sigma^{\text{(cl)}}_{ij}(r')\, \d v(r'),\\
\label{mstress-nl}
&\mu_{ij}(r)=\int_V \alpha(r-r')\,\mu^{\text{(cl)}}_{ij}(r')\, \d v(r').
\end{align}
Using the nonlocal kernel~(\ref{green}) and applying the Helmholtz operator
to (\ref{fstress-nl}) and (\ref{mstress-nl}), one obtains
\begin{align}
\label{fstress-HE-nl}
&\big(1-\varepsilon^{2}\Delta\big)\sigma_{ij}=\sigma^{\text{(cl)}}_{ij},\\
\label{mstress-HE-nl}
&\big(1-\varepsilon^{2}\Delta\big)\mu_{ij}=\mu^{\text{(cl)}}_{ij},
\end{align}
where $\sigma^{\text{(cl)}}_{ij}$ and $\mu^{\text{(cl)}}_{ij}$ are the 
classical local stress and couple stress tensors 
given by
\begin{align}
\label{CE1-nl}
&\sigma^{\text{(cl)}}_{ij}=\lambda\,\delta_{ij}\gamma^{\text{(cl)}}_{kk}
        +(\mu+\eta)\gamma^{\text{(cl)}}_{ij}+(\mu-\eta)\gamma^{\text{(cl)}}_{ji},\\
\label{CE2-nl}
&\mu^{\text{(cl)}}_{ij}=\alpha\,\delta_{ij}\kappa^{\text{(cl)}}_{kk}
+(\beta+\gamma)\kappa^{\text{(cl)}}_{ij}+(\beta-\gamma)\kappa^{\text{(cl)}}_{ji}.
\end{align}
Here $\gamma^{\text{(cl)}}_{ij}$ and $\gamma^{\text{(cl)}}_{ij}$ are the local
micropolar distortion and micropolar bend-twist tensors, respectively.
It is important to note that the local micropolar material constants 
in Eqs.~(\ref{CE1-nl}) and (\ref{CE2-nl}) 
are the same which appear in (\ref{CE1}) and (\ref{CE2}).

Again if we identify $\sigma^{\text{(cl)}}_{ij}=\tl\sigma_{ij}$ and
$\mu^{\text{(cl)}}_{ij}=\tl\mu_{ij}$,
Eqs.~(\ref{fstress-HE}) and (\ref{mstress-HE}) coincide
with (\ref{fstress-HE-nl}) and (\ref{mstress-HE-nl}), respectively. 
In addition the stresses
fulfill the equilibrium conditions~(\ref{FE-NL}) and (\ref{ME-NL})
in gradient micropolar elasticity, too. 
No double force and double couple stresses
appear in nonlocal micropolar elasticity.

\subsection{Screw Dislocation}
The expressions for the force stresses and couple stresses of a screw 
dislocations in a Cosserat  continuum or micropolar medium 
were given by Kessel~\cite{Kessel70} (see also~\cite{Nowacki74,Nowacki86,Minagawa77,Povstenko94}).
In micropolar elasticity,
the asymmetric force stresses of a straight screw dislocation 
may be given in terms of the stress function 
\begin{align}
\label{sf-an-cl}
\tl\sigma {}_{z\varphi}=\pd_r \tl F {}_- \, ,\qquad\qquad
\tl\sigma {}_{\varphi z}=\pd_r \tl F {}_+ \,,
\end{align}
where the two stress functions $\tl F {}_\pm$  are given by
\begin{align}
\label{SF-cl-cc}
\tl F {}_\pm=\frac{b_z}{2\pi}\Big\{\mu\,\ln r\pm \eta\, K_0(r/h)\Big\}\, .
\end{align}
They fulfill
\begin{align}
\big(1-h^2\Delta\big)\Delta \tl F {}_\pm =b_z\big\{
(\mu\mp\eta)[1-h^2\Delta]\pm \eta\big\} \delta(x)\delta(y)\, .
\end{align}

In gradient micropolar elasticity,
the asymmetric force stress can be expressed in terms of new stress
functions $F_-$ and $F_+$:
\begin{align}
\label{sf-an-cc}
\sigma {}_{z\varphi}=\pd_r F {}_- \, ,\qquad\qquad
\sigma {}_{\varphi z}=\pd_r F {}_+ \, .
\end{align}
If we substitute~(\ref{sf-an-cl}), (\ref{SF-cl-cc}) and (\ref{sf-an-cc}) 
into the Helmholtz equation  for the force stress tensor~(\ref{fstress-HE}), 
we obtain for the modified stress functions the following inhomogeneous
Helmholtz equation
\begin{align}
\label{f-fe-cc}
\big(1-\varepsilon^{2}\Delta\big)F_\pm=\frac{b_z}{2\pi}\big\{\mu\,\ln r\pm \eta\, K_0(r/h)\big\}\, ,
\end{align}
where the inhomogeneous part is given by the stress functions~(\ref{SF-cl-cc}). 
The nonsingular solution of~(\ref{f-fe-cc}) is~\cite{LM04a}
\begin{align}
\label{SF-1-cc}
F_\pm=\frac{b_z}{2\pi}\Big\{\mu\big[\ln r+K_0(r/\varepsilon)\big]
\pm \eta\,\frac{h^2}{h^2-\varepsilon^2}\big[ K_0(r/h)-K_0(r/\varepsilon)\big]\Big\}.
\end{align}
It is a more complicated superposition of the micropolar stress function~(\ref{SF-cl-cc})
and a gradient coefficient depending term that in strain gradient elasticity~\footnote{In~\cite{LM04a,LM04b}
we used the notation $\kappa=1/\varepsilon$, $\tau=1/h$ and $\zeta=1/l$. 
In the current notation the pre-factors and the limits to micropolar elasticity
are easier to handle.}. 
In addition, they satisfy
\begin{align}
\big(1-\varepsilon^2\Delta\big) \big(1-h^2\Delta\big)\Delta F {}_\pm =b_z\big\{
(\mu\mp\eta)[1-h^2\Delta]\pm \eta\big\} \delta(x)\delta(y)\, .
\end{align}

Using (\ref{sf-an-cc}) and (\ref{SF-1-cc}), 
the asymmetric force stress is found as
\begin{align}
\label{stress_zp-cc}
&\sigma_{z\varphi}=\frac{b_z}{2\pi}\,\frac{1}{r}
\Big\{\mu\Big[1-\frac{r}{\varepsilon}\, K_1(r/\varepsilon)\Big]
+\frac{\eta h^2}{h^2-\varepsilon^2}\Big[\frac{r}{h}\, K_1(r/h)-\frac{r}{\varepsilon}\, K_1(r/\varepsilon)\Big]\Big\},\\
\label{stress_pz-cc}
&\sigma_{\varphi z}=\frac{b_z}{2\pi}\,\frac{1}{r}
\Big\{\mu\Big[1-\frac{r}{\varepsilon}\, K_1(r/\varepsilon)\Big]
-\frac{\eta h^2}{h^2-\varepsilon^2}\Big[\frac{r}{h}\, K_1(r/h)-\frac{r}{\varepsilon}\, K_1(r/\varepsilon)\Big]\Big\}.
\end{align}
\begin{figure}[t]\unitlength1cm
\centerline{
(a)
\begin{picture}(8,6)
\put(0.0,0.2){\epsfig{file=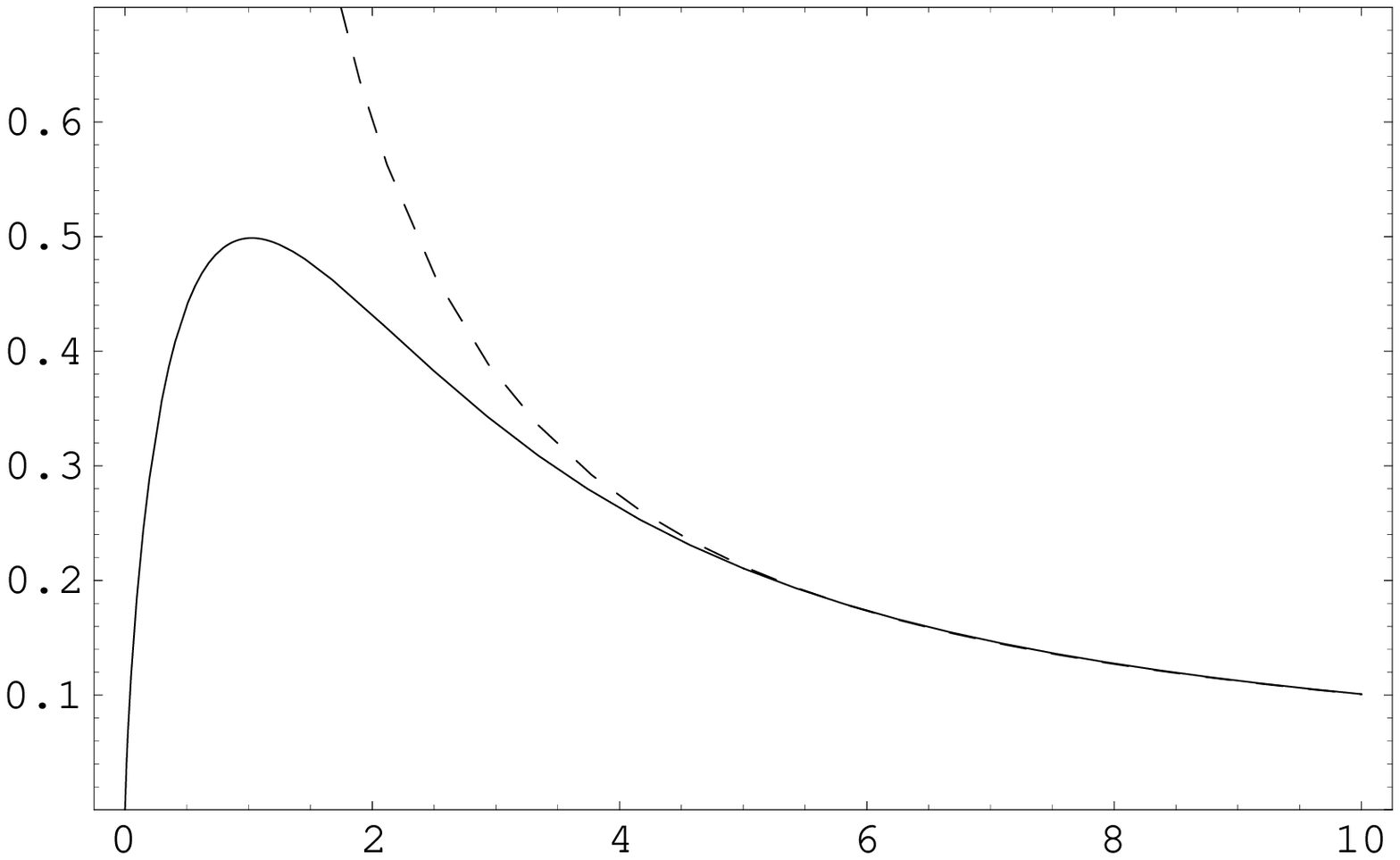,width=9cm}}
\put(4.5,0.0){$r/\varepsilon$}
\put(-1.0,4.5){$\sigma_{z\varphi}$}
\end{picture}
}
\centerline{
(b)
\begin{picture}(8,6)
\put(0.0,0.2){\epsfig{file=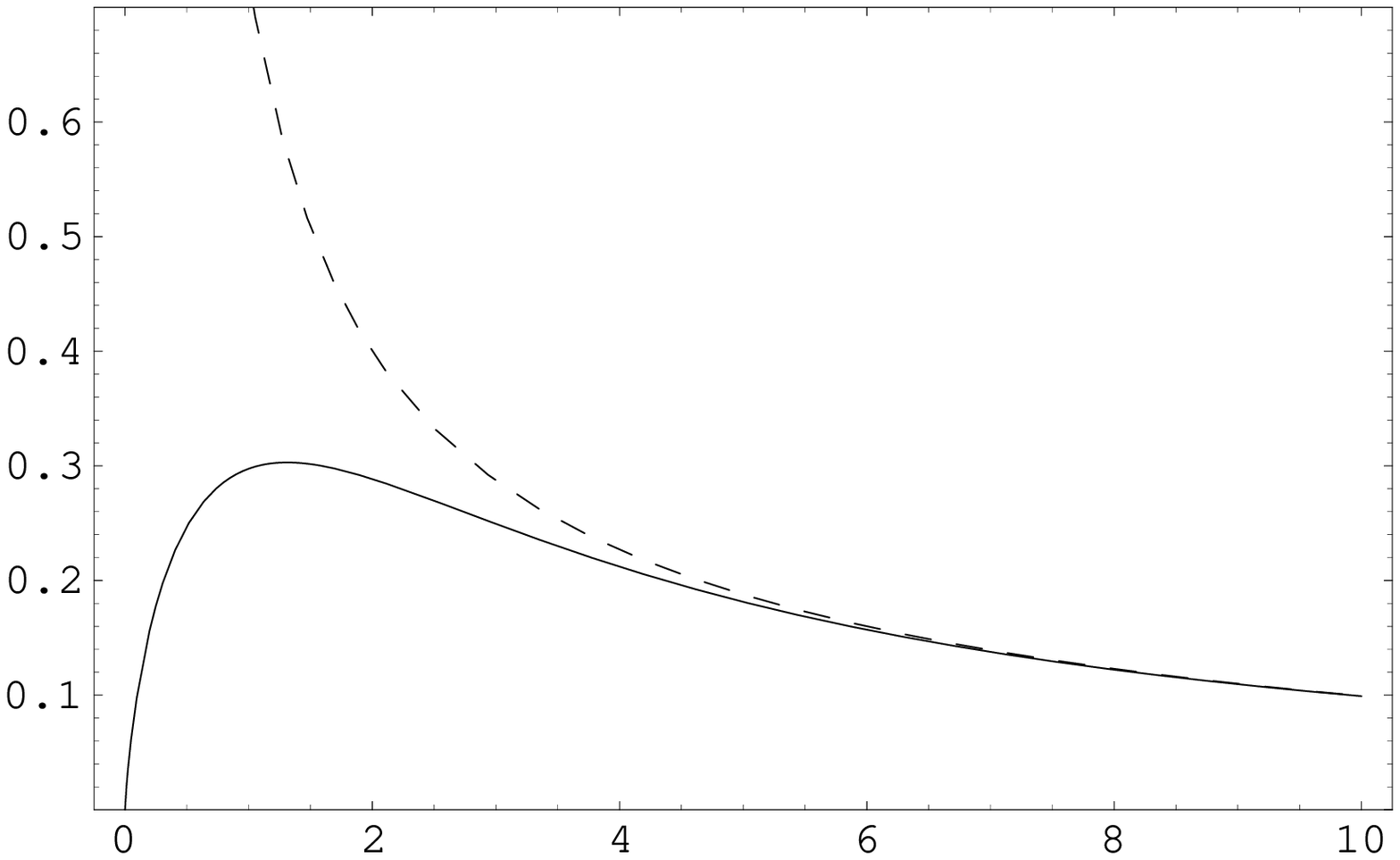,width=9cm}}
\put(4.5,0.0){$r/\varepsilon$}
\put(-1.0,4.5){$\sigma_{\varphi z}$}
\end{picture}
}
\caption{The components of the asymmetric stress of a screw dislocation:
(a)~$\sigma_{z\varphi}$ and (b)~$\sigma_{\varphi z}$ 
are given in units of $\mu b_z/[2\pi\varepsilon]$ with $h=2\varepsilon$
and $\mu=3\eta$.  
The dashed curves represent the micropolar results.}
\label{fig:T-screw-CC}
\end{figure}
They are zero at $r=0$ and have extremum values near the dislocation line. 
The extremum values depend on $h$ and $\varepsilon$.
For instance, with $h=2\varepsilon$ and $\mu=3\eta$ we obtain:
$\sigma_{z\varphi}\simeq 0.499\, \mu b_z/[2\pi\varepsilon]$ at $r\simeq 1.024\, \varepsilon$ 
and
$\sigma_{\varphi z}\simeq 0.303\, \mu b_z/[2\pi\varepsilon]$ at $r\simeq 1.312\, \varepsilon$. 
Thus, $\sigma_{z\varphi}>\sigma_{\varphi z}$.
The stresses are plotted versus $r/\varepsilon$ in Fig.~\ref{fig:T-screw-CC}.
We note that the stresses~(\ref{stress_zp-cc}) and (\ref{stress_pz-cc}) are also 
solutions of Eq.~(\ref{FE-NL}) in nonlocal micropolar elasticity. 
In nonlocal micropolar elasticity
these solutions were calculated by Povstenko~\cite{Povstenko98}
(He used the transposed tensors.).
In addition, in the limit to micropolar elasticity, $\varepsilon\rightarrow 0$,
we recover in (\ref{stress_zp-cc}) and (\ref{stress_pz-cc}) Kessel's
result~\cite{Kessel70}.

By means of the inverse of the constitutive relation~(\ref{CE1}), 
the micropolar distortion is obtained as
\begin{align}
\label{dist-zp-CC}
&\gamma_{z\varphi}=\frac{b_z}{4\pi}\,\frac{1}{r}
\Big\{\Big[1-\frac{r}{\varepsilon}\, K_1(r/\varepsilon)\Big]
+\frac{h^2}{h^2-\varepsilon^2}\Big[\frac{r}{h}\, K_1(r/h)-\frac{r}{\varepsilon}\, K_1(r/\varepsilon)\Big]\Big\},\\
\label{dist-pz-CC}
&\gamma_{\varphi z}=\frac{b_z}{4\pi}\,\frac{1}{r}
\Big\{\Big[1-\frac{r}{\varepsilon}\, K_1(r/\varepsilon)\Big]
-\frac{h^2}{h^2-\varepsilon^2}\Big[\frac{r}{h}\, K_1(r/h)-\frac{r}{\varepsilon}\, K_1(r/\varepsilon)\Big]\Big\}.
\end{align}
\begin{figure}[t]\unitlength1cm
\centerline{
(a)
\begin{picture}(8,6)
\put(0.0,0.2){\epsfig{file=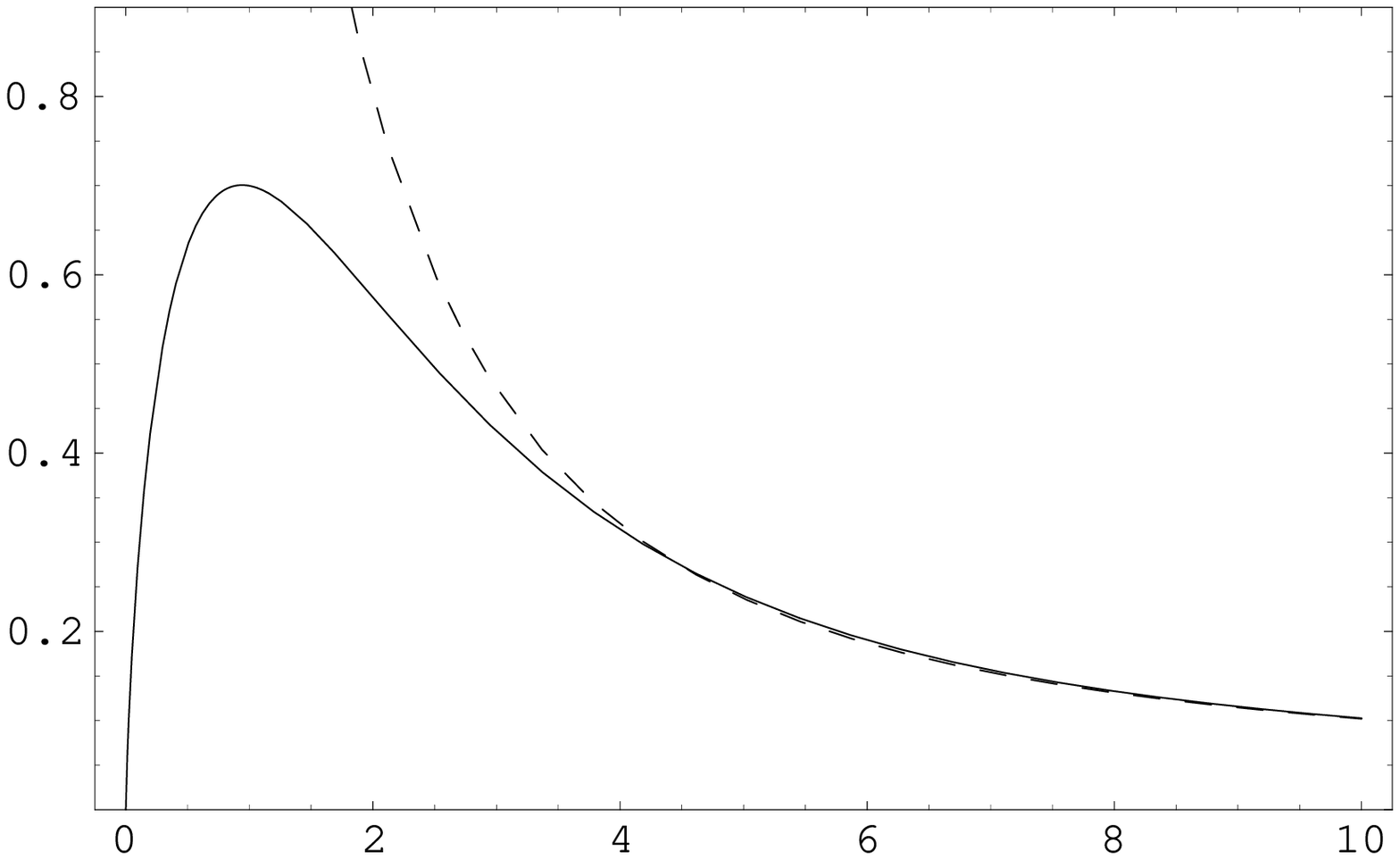,width=9cm}}
\put(4.5,0.0){$r/\varepsilon$}
\put(-1.0,4.5){$\gamma_{z\varphi}$}
\end{picture}
}
\centerline{
(b)
\begin{picture}(8,6)
\put(0.0,0.2){\epsfig{file=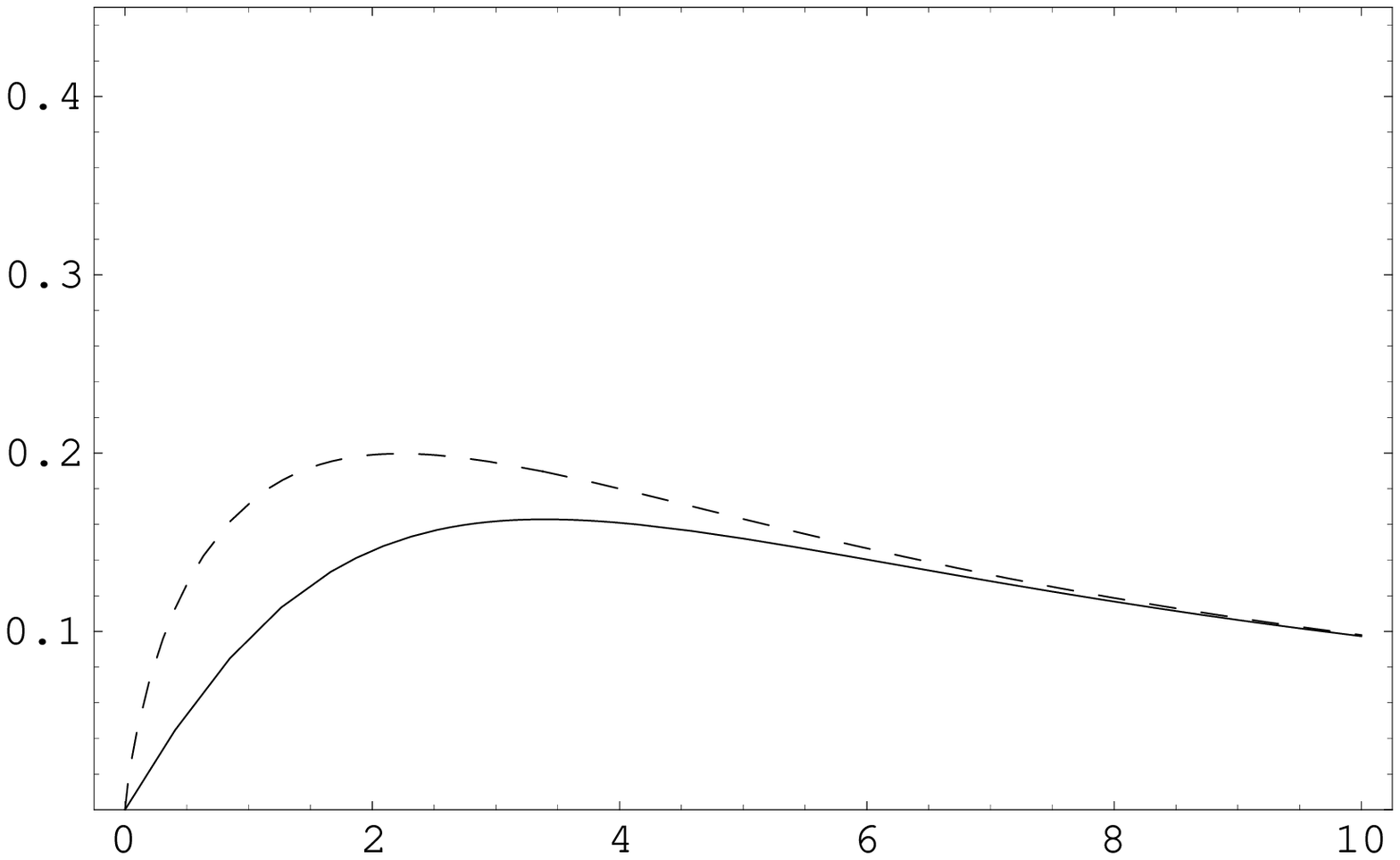,width=9cm}}
\put(4.5,0.0){$r/\varepsilon$}
\put(-1.0,4.5){$\gamma_{\varphi z}$}
\end{picture}
}
\caption{The components of the micropolar distortion for a screw dislocation:
(a)~$\gamma_{z\varphi}$ and (b)~$\gamma_{\varphi z}$ 
are given in units of $ b_z/[4\pi\varepsilon]$ and $h=2\varepsilon$.  
The dashed curves represent the micropolar results.}
\label{fig:dist-CC}
\end{figure}
They are zero at $r=0$ and have extremum values near the dislocation line. 
Again, the extremum values depend on $h$ and $\varepsilon$.
For instance, with $h=2\varepsilon$ we have:
$\gamma_{z\varphi}\simeq 0.701\, b_z/[4\pi\varepsilon]$ at $r\simeq 0.941\, \varepsilon$ 
and
$\gamma_{\varphi z}\simeq 0.163\, b_z/[4\pi\varepsilon]$ at $r\simeq 3.395\, \varepsilon$. 
Thus, it yields $\gamma_{z\varphi}>\gamma_{\varphi z}$.
In addition, $\tl\gamma_{\varphi z}\simeq 0.151\, b_z/[4\pi\varepsilon]$ 
at $r\simeq 2.229\, \varepsilon$. The micropolar distortions~(\ref{dist-zp-CC}) and
(\ref{dist-pz-CC}) are plotted in Fig.~\ref{fig:dist-CC} versus $r/\varepsilon$ and with 
$h=2\varepsilon$.  
It is interesting to note that the elastic micropolar 
strain $\gamma_{(z\varphi)}$ has the same
form as the elastic strain $E_{z\varphi}$ calculated in gradient elasticity 
(see Eq.~\ref{E-cyl}).
In the limit $\varepsilon\rightarrow 0$,
we recover in (\ref{dist-zp-CC}) and (\ref{dist-pz-CC}) 
Minagawa's result~\cite{Minagawa77}.

The micro-rotation of a straight screw dislocation reads~\cite{LM04a}
\begin{align}
\label{mr-scr}
\varphi_r=\gamma_{\varphi z}=\frac{1}{2}\, \pd_r F_+(\mu=1,\eta=1)\, .
\end{align}
On the other hand, $\varphi_r$ follows directly from Eq.~(\ref{rot-HE})
with $\tl\varphi_r=\tl\gamma_{\varphi z}$.
The micropolar bend-twist or wryness is given in terms of~(\ref{mr-scr})
\begin{align}
\label{mbt-scr}
\kappa_{rr}=\pd_r \varphi_r\, ,\qquad
\kappa_{\varphi\varphi}=\frac{1}{r}\, \varphi_r\, .
\end{align}
So, we find
\begin{align}
\label{k-rr-CC}
&\kappa_{rr}=-\frac{b_z}{4\pi}\frac{1}{r^2}\Big\{
1-\frac{1}{h^2-\varepsilon^2}
\Big[h r\, K_1(r/h)-\varepsilon r\, K_1(r/\varepsilon)
+r^2 \big(K_0(r/h)-K_0(r/\varepsilon)\big)\Big]\Big\},\\
\label{k-vv-CC}
&\kappa_{\varphi\varphi}=\frac{b_z}{4\pi}\frac{1}{r^2}\Big\{
1-\frac{1}{h^2-\varepsilon^2}
\Big[h r\, K_1(r/h)-\varepsilon r\, K_1(r/\varepsilon)\Big]\Big\}.
\end{align}  
The trace of the micropolar bend-twist tensor reads
\begin{align}
&\kappa_{kk}=\frac{b_z}{4\pi}\,\frac{1}{h^2-\varepsilon^2}
\Big[K_0(r/h)-K_0(r/\varepsilon)\Big].
\end{align}
The components $\kappa_{rr}$, $\kappa_{\varphi\varphi}$ and
$\kappa_{kk}$ have a maximum at the dislocation line.
In fact, 
one obtains at $r=0$:
\begin{align}
\kappa_{rr}=\kappa_{\varphi\varphi}=\frac{1}{2}\kappa_{kk}\simeq
\frac{b_z}{4\pi}\,\frac{1}{2(h^2-\varepsilon^2)}\, \ln\frac{h}{\varepsilon}\,.
\end{align}
For the choice $h=2\varepsilon$ we have at $r=0$:
\begin{align}
\kappa_{rr}=\kappa_{\varphi\varphi}=\frac{1}{2}\kappa_{kk}\simeq
0.116\,\frac{b_z}{4\pi\varepsilon^2}\, .
\end{align}
Thus, all $1/r^2$- and $\ln r$-singularities, which are present
in the results of Cosserat theory, are eliminated.
In addition, the bend-twist is modified in the region $r/\varepsilon <6$. 
The shape of the bend-twist in gradient micropolar elasticity looks more
physical than the bend-twist in classical micropolar elasticity 
(see Fig.~\ref{fig:curv-CC}). 
\begin{figure}[p]\unitlength1cm
\vspace*{-1.0cm}
\centerline{
(a)
\begin{picture}(8,6)
\put(0.0,0.2){\epsfig{file=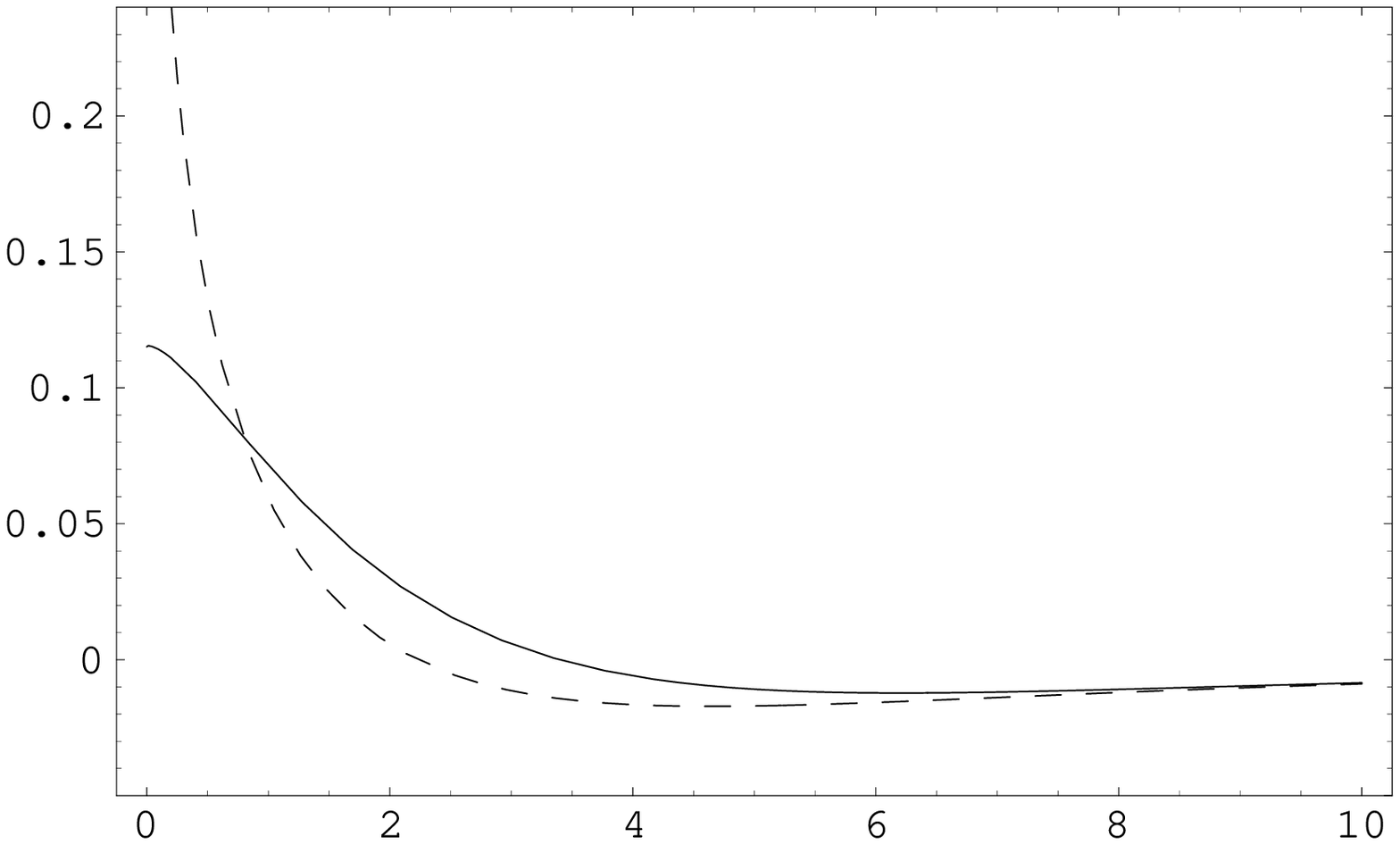,width=9cm}}
\put(4.5,0.0){$r/\varepsilon$}
\put(-1.0,4.5){$\kappa_{rr}$}
\end{picture}
}
\centerline{
(b)
\begin{picture}(8,6)
\put(0.0,0.2){\epsfig{file=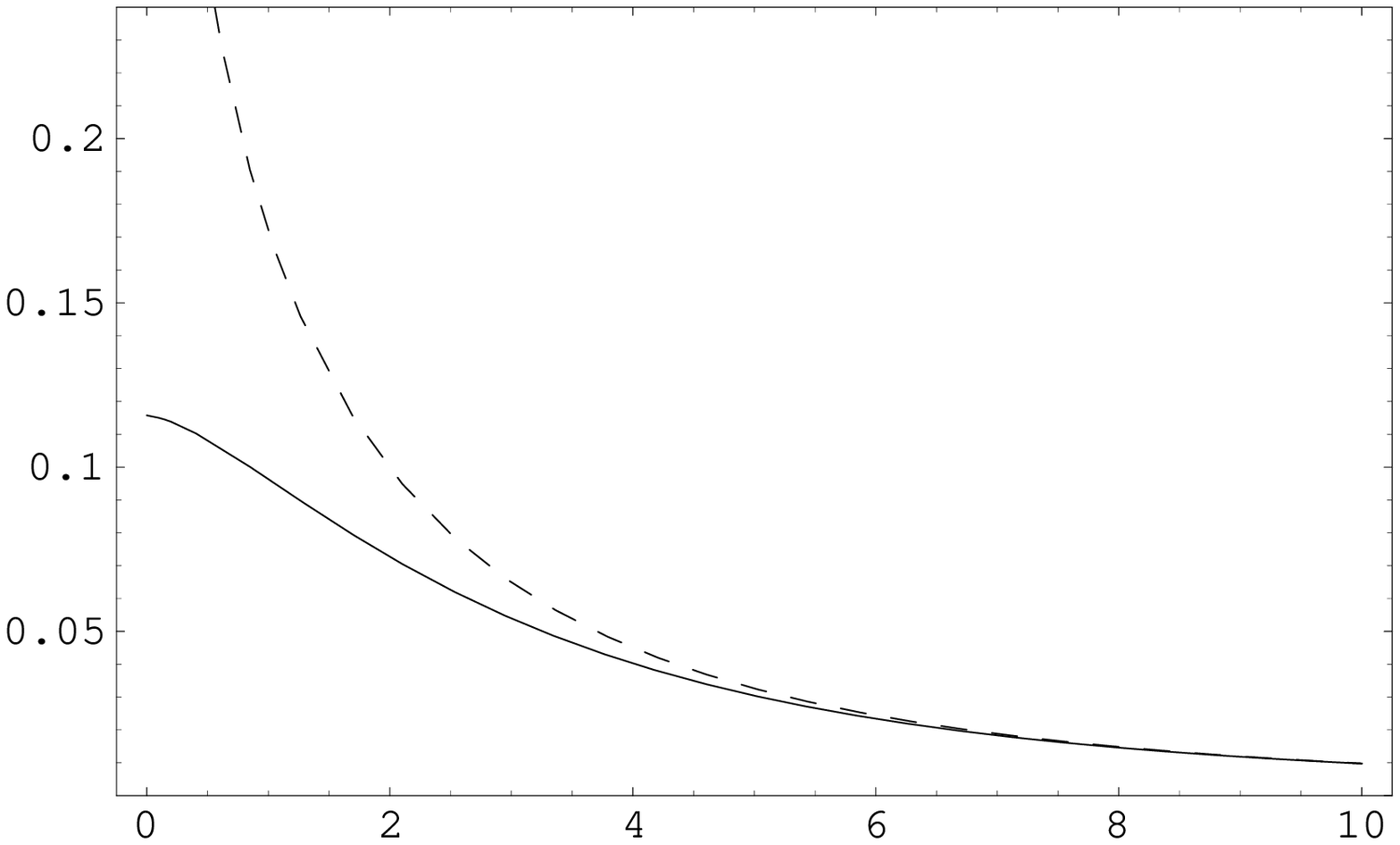,width=9cm}}
\put(4.5,0.0){$r/\varepsilon$}
\put(-1.0,4.5){$\kappa_{\varphi\varphi}$}
\end{picture}
}
\centerline{
(c)
\begin{picture}(8,6)
\put(0.0,0.2){\epsfig{file=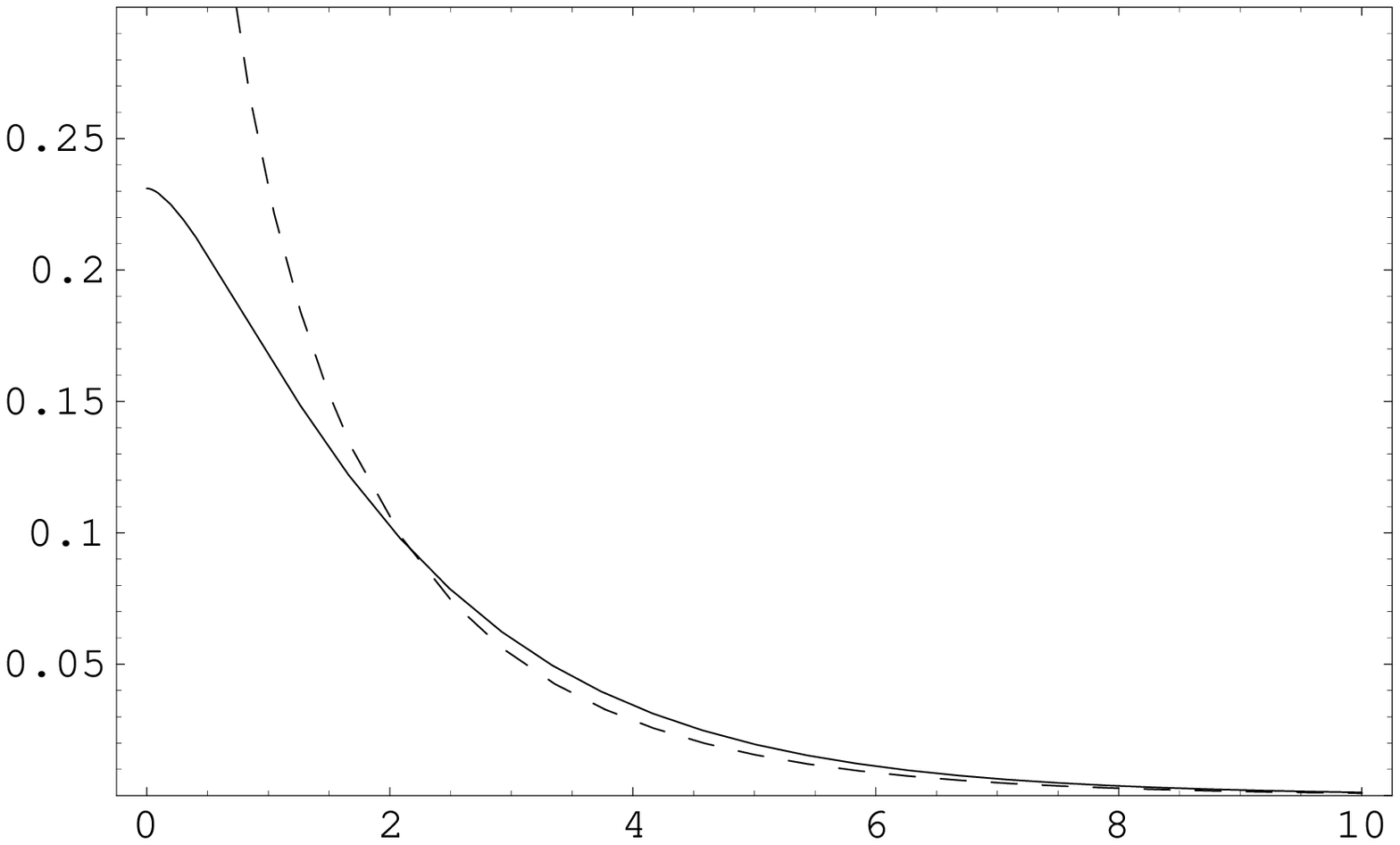,width=9cm}}
\put(4.5,0.0){$r/\varepsilon$}
\put(-1.0,4.4){$\kappa_{kk}$}
\end{picture}
}
\caption{The components of the micropolar bend-twist tensor for a screw dislocation:
(a)~$\kappa_{rr}$, (b)~$\kappa_{\varphi\varphi}$ and (c)~$\kappa_{kk}$ 
are given in units of $b_z/[4\pi\varepsilon^2]$ and with $h=2\varepsilon$.  
The dashed curves represent the classical (micropolar) components.}
\label{fig:curv-CC}
\end{figure}
In addition, it can be seen that $\kappa_{rr}$ is negative in the far field expression
and positive near the dislocation line. The point at which it changes 
sign is for $h=2\varepsilon$: $r\simeq 3.395\, \varepsilon$ (see Fig.~\ref{fig:curv-CC}a).
In the limit $\varepsilon\rightarrow 0$,
we recover in (\ref{k-rr-CC}) and (\ref{k-vv-CC}) 
Minagawa's result~\cite{Minagawa77}.

Using Eq.~({\ref{CE2}) one obtains 
from the micropolar bend-twist tensor the couple-stress tensor.
The non-vanishing components of the micropolar couple stress tensor
are
\begin{align}
\label{mu-rr-screw}
&\mu_{rr}=-\frac{b_z}{2\pi}\frac{1}{r^2}\Big\{
\beta-\frac{1}{h^2-\varepsilon^2}
\Big[\beta\big(h r\, K_1(r/h)-\varepsilon r\, K_1(r/\varepsilon)\big)
+\frac{\alpha+2\beta}{2}\,r^2 \big(K_0(r/h)-K_0(r/\varepsilon)\big)\Big]\Big\},\\
\label{mu-vv-screw}
&\mu_{\varphi\varphi}=\frac{b_z}{2\pi}\frac{1}{r^2}\Big\{
\beta-\frac{1}{h^2-\varepsilon^2}
\Big[\beta\big( h r\, K_1(r/h)-\varepsilon r\, K_1(r/\varepsilon)\big)
-\frac{\alpha}{2}\, \big(K_0(r/h)-K_0(r/\varepsilon)\big)
\Big]\Big\},\\
\label{mu-zz-screw}
&\mu_{zz}=\frac{b_z\alpha}{4\pi}\,\frac{1}{h^2-\varepsilon^2}
\Big[K_0(r/h)-K_0(r/\varepsilon)\Big]\, .
\end{align}
The trace of the couple stress tensor is given by
\begin{align}
\mu_{kk}=\frac{b_z(3\alpha+2\beta)}{4\pi}\,\frac{1}{h^2-\varepsilon^2}
\Big[K_0(r/h)-K_0(r/\varepsilon)\Big]\, .
\end{align}
All components of the couple stress tensor are nonsingular in the gradient
micropolar elasticity.
In fact, 
one obtains at $r=0$:
\begin{align}
\mu_{zz}=
\frac{\alpha}{\beta+\alpha}\,
\mu_{\varphi\varphi}=
\frac{\alpha}{\beta+\alpha}\,
\mu_{rr}=
\frac{\alpha}{2\beta+3\alpha}\,
\mu_{kk}\simeq
\alpha\,\frac{b_z}{2\pi}\,\frac{1}{2(h^2-\varepsilon^2)}\,\ln\frac{h}{\varepsilon}\, .
\end{align}
The micropolar bend-twist and the couple stress tensors 
of a screw dislocation are symmetric tensors.
The far-fields of~(\ref{k-rr-CC}) and (\ref{k-vv-CC}), and
(\ref{mu-rr-screw})--(\ref{mu-zz-screw}) agree 
with the result given by Kessel~\cite{Kessel70} and Minagawa~\cite{Minagawa77}.
In the limit $\varepsilon\rightarrow 0$,
we recover in (\ref{mu-rr-screw})--(\ref{mu-zz-screw}) 
Kessel's result~\cite{Kessel70}.

The couple stresses~(\ref{mu-rr-screw})--(\ref{mu-zz-screw}) 
are the solutions in nonlocal elasticity.
They fulfill together with 
the stresses~(\ref{stress_zp-cc}) and (\ref{stress_pz-cc}) the
moment equilibrium~(\ref{FE-NL}) in nonlocal micropolar elasticity. 
Using the relation $K_2(x)-K_0(x)=2/x\, K_1(x)$, 
it can be shown that the couple stresses  agree with the 
couple stresses in nonlocal micropolar elasticity
calculated by Povstenko~\cite{Povstenko98} 
up to a pre-factor $1/2$ which is missing in his formulas.
This is because he used Nowacki's expressions~\cite{Nowacki74} for the couples stresses 
of a screw dislocation in Cosserat theory which are the source of
this missing factor $1/2$. The results of Kessel~\cite{Kessel70} and
Minagawa~\cite{Minagawa77} have the correct pre-factor. 
The components of the force and couple stress, distortion and bend-twist tensors 
in Cartesian coordinates are given in Ref.~\cite{LM04a}.

\subsection{Edge Dislocation}
The expressions of the force stresses and couple stresses 
of a straight edge dislocations in micropolar elasticity (Cosserat theory) 
are given by~\cite{Kessel70,Knesl,Nowacki74,Nowacki86,Minagawa79}.
We use the stress function formulation
given by~\cite{Mindlin63,Schaefer62} (see also~\cite{Eringen99,Nowacki86,Povstenko94}),
which involves two stress functions $\tl f$ 
and $\tl \Psi$, which are connected with the
force stresses and couple stresses by
\begin{align}
&\tl\sigma_{rr}=\frac{1}{r}\,\pd_r\tl f+\frac{1}{r^2}\,\pd^2_{\varphi\varphi}\tl f
            -\frac{1}{r}\,\pd^2_{r\varphi}\tl \Psi+\frac{1}{r^2}\,\pd_{\varphi}\tl \Psi,\\
&\tl\sigma_{\varphi\varphi}=\pd^2_{rr}\tl f
        +\frac{1}{r}\,\pd^2_{r\varphi}\tl \Psi-\frac{1}{r^2}\,\pd_{\varphi}\tl \Psi,\\
&\tl\sigma_{r\varphi}=-\frac{1}{r}\,\pd^2_{r\varphi}\tl f+\frac{1}{r^2}\,\pd_{\varphi}\tl f
                   +\pd^2_{rr}\tl \Psi,\\\
&\tl\sigma_{\varphi r}=-\frac{1}{r}\,\pd^2_{r\varphi}\tl f+\frac{1}{r^2}\,\pd_{\varphi}\tl f
                   -\frac{1}{r}\,\pd_{r}\tl \Psi-\frac{1}{r^2}\,\pd^2_{\varphi\varphi}\tl \Psi,\\
&\tl\mu_{zr}=\pd_r\tl\Psi,\qquad
\tl\mu_{z\varphi}=\frac{1}{r}\,\pd_\varphi\tl\Psi
\end{align}
and $\tl\sigma_{zz}=\nu(\tl\sigma_{rr}+\tl\sigma_{\varphi\varphi})$.
Here $\tl f$ is the Airy stress function known from classical elasticity 
and $\tl \Psi$ is the stress function introduced by Mindlin~\cite{Mindlin63},
Schaefer~\cite{Schaefer62} and Carlson~\cite{Carlson}.
For a straight edge dislocation the so-called Airy-Mindlin stress functions
reads~\cite{Nowacki74,Nowacki86}
\begin{align}
\label{SF1-ed-cl}
\tl f&=-\frac{\mu b_x}{2\pi(1-\nu)}\, \sin\varphi\,  \big(r \ln r \big)\, ,\\
\label{SF2-ed-cl}
\tl\Psi&=-\frac{(\beta+\gamma)b_x}{2\pi}\,\frac{\cos\varphi}{r}\Big(1-\frac{r}{l}\, K_1(r/l)\Big) 
\end{align}
with
\begin{align}
\label{SF2-HE}
\big(1-l^2\Delta\big)\tl\Psi=-\frac{(\beta+\gamma)b_x}{2\pi}\,\frac{\cos\varphi}{r} .
\end{align}
With the help of Eq.~(\ref{SF2-HE}) and some simple manipulations, 
we find the Cauchy-Riemann relations
\begin{align}
\pd_r\big(1-l^2\Delta\big)\tl\Psi &=-\frac{2(1-\nu)}{\mu(\beta+\gamma)}\,\frac{1}{r}\,\pd_\varphi \big(\Delta \tl f\big),\\
\frac{1}{r}\,\pd_\varphi\big(1-l^2\Delta\big)\tl\Psi &=\frac{2(1-\nu)}{\mu(\beta+\gamma)}\,\pd_r  \big(\Delta \tl f
\big).
\end{align}
By construction this stress function ansatz fulfills
the equilibrium conditions~(\ref{FE-NL}) and (\ref{ME-NL}).
All components of the force stress tensor have  $1/r$-singularities and 
those of the couple stress tensor have $1/r^2$-singularities.

In gradient micropolar elasticity we make the following ansatz
\begin{align}
&\sigma_{rr}=\frac{1}{r}\,\pd_r f+\frac{1}{r^2}\,\pd^2_{\varphi\varphi} f
            -\frac{1}{r}\,\pd^2_{r\varphi} \Psi+\frac{1}{r^2}\,\pd_{\varphi} \Psi,\\
&\sigma_{\varphi\varphi}=\pd^2_{rr} f
+\frac{1}{r}\,\pd^2_{r\varphi} \Psi-\frac{1}{r^2}\,\pd_{\varphi} \Psi,\\
&\sigma_{r\varphi}=-\frac{1}{r}\,\pd^2_{r\varphi} f+\frac{1}{r^2}\,\pd_{\varphi} f
                   +\pd^2_{rr} \Psi,\\\
&\sigma_{\varphi r}=-\frac{1}{r}\,\pd^2_{r\varphi} f+\frac{1}{r^2}\,\pd_{\varphi} f
                   -\frac{1}{r}\,\pd_{r} \Psi-\frac{1}{r^2}\,\pd^2_{\varphi\varphi} \Psi,\\
\label{mu-SFA}
&\mu_{zr}=\pd_r\Psi,\qquad
\mu_{z\varphi}=\frac{1}{r}\,\pd_\varphi\Psi
\end{align}
and $\sigma_{zz}=\nu(\sigma_{rr}+\sigma_{\varphi\varphi})$.
It has the same form as the stress function ansatz 
in micropolar elasticity. 
Here $f$ and $\Psi$ are the modified stress functions 
which must be calculated.
So, we obtain for the modified stress functions the following inhomogeneous
Helmholtz equations 
\begin{align}
\label{f_fe-cc}
&\big(1-\varepsilon^{2}\Delta\big)f=
-\frac{\mu b_x}{2\pi(1-\nu)}\, \sin\varphi\,  \big(r \ln r \big)\, ,\\
\label{psi_fe-cc}
&\big(1-\varepsilon^{2}\Delta\big)\Psi=
-\frac{(\beta+\gamma)b_x}{2\pi}\,\frac{\cos\varphi}{r}\Big(1-\frac{r}{l}\, K_1(r/l)\Big)\, .
\end{align}
The nonsingular solutions are (see~\cite{LM04a,LM04b} for technical details) 
\begin{align}
\label{SF1-ed}
&f=-\frac{\mu b_x}{2\pi(1-\nu)}\, \sin\varphi \Big\{r \ln r 
+\frac{2\varepsilon^2}{r}\Big(1-\frac{r}{\varepsilon}\, K_1(r/\varepsilon)\Big)\Big\},\\
\label{SF2-ed}
&\Psi=-\frac{(\beta+\gamma)b_x}{2\pi}\,\frac{\cos\varphi}{r}
\Big\{1-\frac{1}{l^2-\varepsilon^2}
\Big[\frac{r}{l}\, K_1(r/l)-\frac{r}{\varepsilon}\, K_1(r/\varepsilon)\Big]\Big\}.
\end{align}
Again, they are superpositions of the classical stress functions~(\ref{SF1-ed-cl})
and (\ref{SF2-ed-cl}) and the gradient parts which depend on the 
gradient coefficient $\varepsilon$.
Therefore, the stress function~(\ref{SF1-ed}) has the same form as in
strain gradient elasticity (see Eq.~(\ref{SFA-edge})).
The stress function~(\ref{SF2-ed}) also satisfies the 
following bi-Helmholtz equation:
\begin{align}
\big(1-l^2\Delta\big)\big(1-\varepsilon^{2}\Delta\big)\Psi=
-(\beta+\gamma)b_x\,\pd_x \delta(x)\delta(y)\, .
\end{align}
Thus, it is the corresponding Green function.
In addition, the stress functions $\Psi$ and $f$ are related by 
the following Cauchy-Riemann relations 
\begin{align}
\pd_r\big(1-\varepsilon^2\Delta\big)
\big(1-l^2\Delta\big)\Psi &=
-\frac{2(1-\nu)}{\mu(\beta+\gamma)}\,\frac{1}{r}\,\pd_\varphi 
\big(1-\varepsilon^2\Delta\big) \Delta f ,\\
\frac{1}{r}\,\pd_\varphi
\big(1-\varepsilon^2\Delta\big)
\big(1-l^2\Delta\big)\Psi &=\frac{2(1-\nu)}{\mu(\beta+\gamma)}\,\pd_r  
\big(1-\varepsilon^2\Delta\big) \Delta f .
\end{align}

Using the Airy-Mindlin stress functions~(\ref{SF1-ed}) and (\ref{SF2-ed}), 
we find for the asymmetric force stress of a straight edge dislocation
\begin{align}
\label{T_rr-cc}
\sigma_{rr}&=-\frac{\mu b_x}{2\pi(1-\nu)}\, 
\frac{\sin\varphi}{r}\Big\{1-\frac{4\varepsilon^2}{r^2}+2 K_2(r/\varepsilon)\Big\}
\nonumber\\
&\quad
+\frac{(\beta+\gamma)b_x}{2\pi}\, \frac{\sin\varphi}{r^3}
\Big\{2-\frac{r^2}{l^2-\varepsilon^2}
\big[K_2(r/l)-K_2(r/\varepsilon)\big]\Big\},\\
\label{T_pp-cc}
\sigma_{\varphi\varphi}&=-\frac{\mu b_x}{2\pi(1-\nu)}\, 
\frac{\sin\varphi}{r}\Big\{1+\frac{4\varepsilon^2}{r^2}
-2 K_2(r/\varepsilon)-2\,\frac{r}{\varepsilon}\, K_1(r/\varepsilon)\Big\}\nonumber\\
&\quad-\frac{(\beta+\gamma)b_x}{2\pi}\, \frac{\sin\varphi}{r^3}
\Big\{2-\frac{r^2}{l^2-\varepsilon^2}
\big[K_2(r/l)-K_2(r/\varepsilon)\big]\Big\},\\
\label{T_rp-cc}
\sigma_{r\varphi}&=\frac{\mu b_x}{2\pi(1-\nu)}\, 
\frac{\cos\varphi}{r}\Big\{1-\frac{4\varepsilon^2}{r^2}+2 K_2(r/\varepsilon)\Big\}\nonumber\\
&\quad-\frac{(\beta+\gamma)b_x}{2\pi}\, \frac{\cos\varphi}{r^3}
\Big\{2-\frac{r^2}{l^2-\varepsilon^2}
\big[K_2(r/l)-K_2(r/\varepsilon)
+\frac{r}{l}\, K_1(r/l)-\frac{r}{\varepsilon}\, K_1(r/\varepsilon)\big]\Big\},\\
\label{T_pr-cc}
\sigma_{\varphi r}&=\frac{\mu b_x}{2\pi(1-\nu)}\, 
\frac{\cos\varphi}{r}\Big\{1-\frac{4\varepsilon^2}{r^2}+2 K_2(r/\varepsilon)\Big\}\nonumber\\
&\quad-\frac{(\beta+\gamma)b_x}{2\pi}\, \frac{\cos\varphi}{r^3}
\Big\{2-\frac{r^2}{l^2-\varepsilon^2}
\big[K_2(r/l)-K_2(r/\varepsilon)\big]\Big\},\\
\label{T_zz-cc}
\sigma_{zz}&=-\frac{\mu b_x\nu }{\pi(1-\nu)}\, 
\frac{\sin\varphi}{r}\Big\{1-\frac{r}{\varepsilon}\, K_1(r/\varepsilon)\Big\},
\end{align}
and
\begin{align}
\label{T_kk-cc}
\sigma_{kk}=-\frac{\mu b_x(1+\nu)}{\pi(1-\nu)}\, 
\frac{\sin\varphi}{r}\Big\{1-\frac{r}{\varepsilon}\, K_1(r/\varepsilon)\Big\}.
\end{align}
The stress fields have no artificial singularities at the dislocation line  
and the extremum stress occurs at a short distance away from 
the dislocation line (see~ Figs.~\ref{fig:T-edge-cc}). 
The stresses~(\ref{T_zz-cc}) and (\ref{T_kk-cc}) are not influenced by the 
Cosserat constants and, thus, they have the same form as the 
corresponding expressions (\ref{T_zz}) and (\ref{T_kk})
in strain gradient elasticity. In Eqs.~(\ref{T_rr-cc})--(\ref{T_pr-cc})
the first parts agree with the stresses~(\ref{T_rr})--(\ref{T_pp}) 
in strain gradient elasticity
and the other parts are modified stresses due to the 
gradient micropolar elasticity. 
Near the dislocation line one can see the difference between the stresses 
calculated in strain gradient elasticity 
and in gradient micropolar elasticity.  
The extremum values of these stresses are changed. 
They are higher or lower than the extremum values in strain gradient theory
(see Fig.~\ref{fig:T-edge-cc}). In addition, the values and the positions of the 
extremum values depend strongly on the material constants of gradient micropolar 
elasticity. 
\begin{figure}[p]\unitlength1cm
\vspace*{-1.0cm}
\centerline{
(a)
\begin{picture}(8,6)
\put(0.0,0.2){\epsfig{file=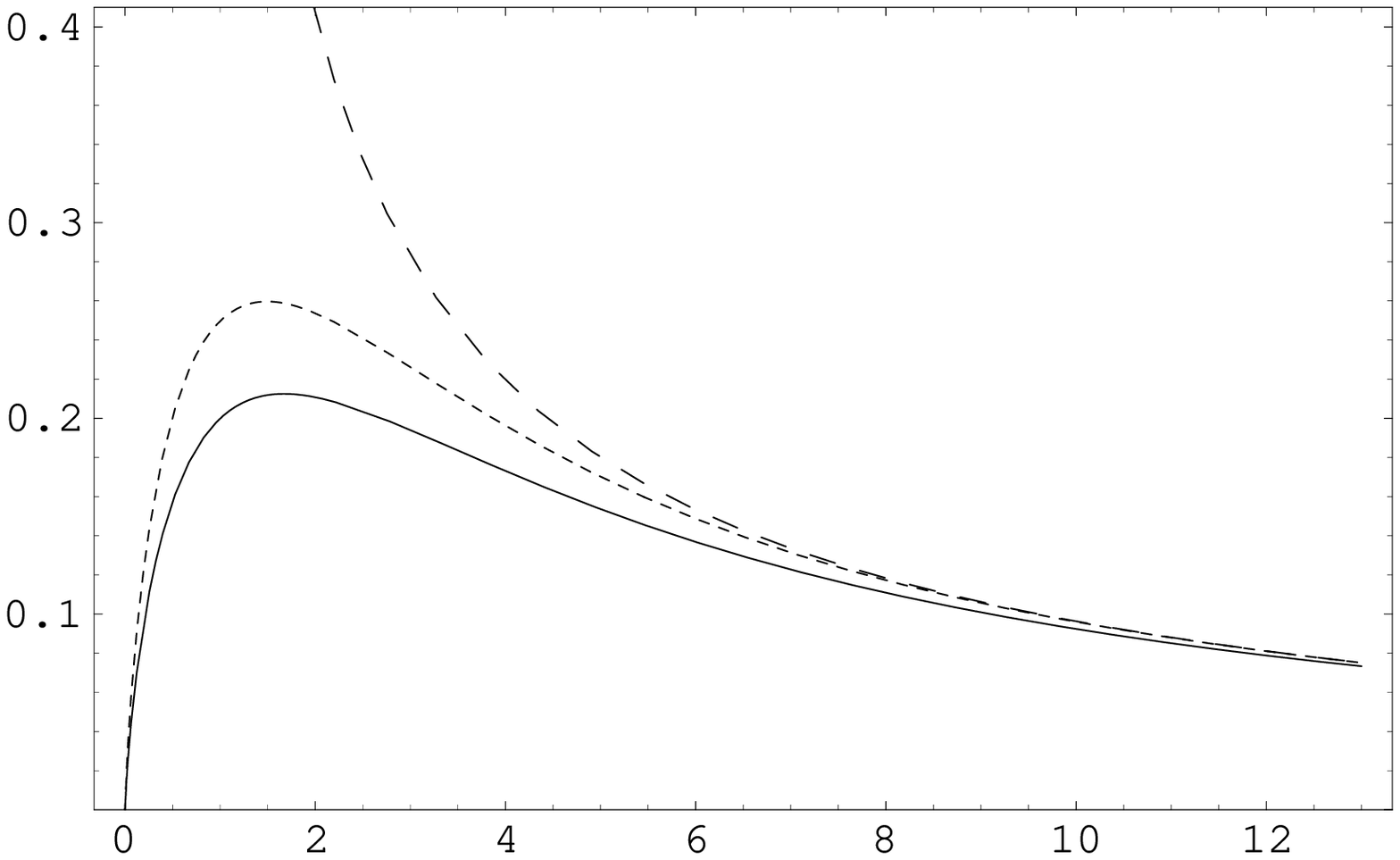,width=9cm}}
\put(4.5,0.0){$r/\varepsilon$}
\put(-1.0,4.5){$\sigma_{rr}$}
\end{picture}
}
\centerline{
(b)
\begin{picture}(8,6)
\put(0.0,0.2){\epsfig{file=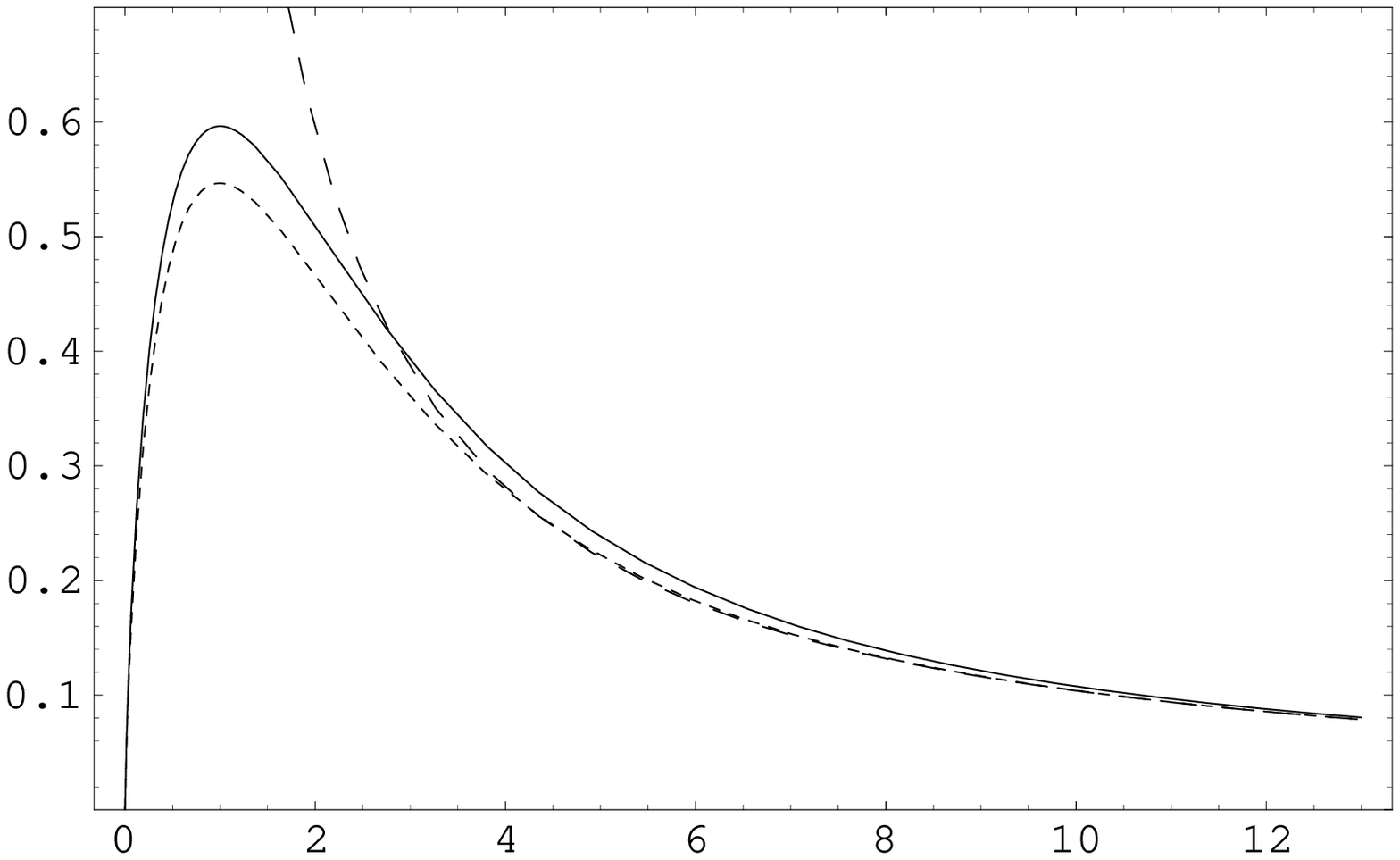,width=9cm}}
\put(4.5,0.0){$r/\varepsilon$}
\put(-1.0,4.5){$\sigma_{\varphi\varphi}$}
\end{picture}
}
\centerline{
(c)
\begin{picture}(8,6)
\put(0.0,0.2){\epsfig{file=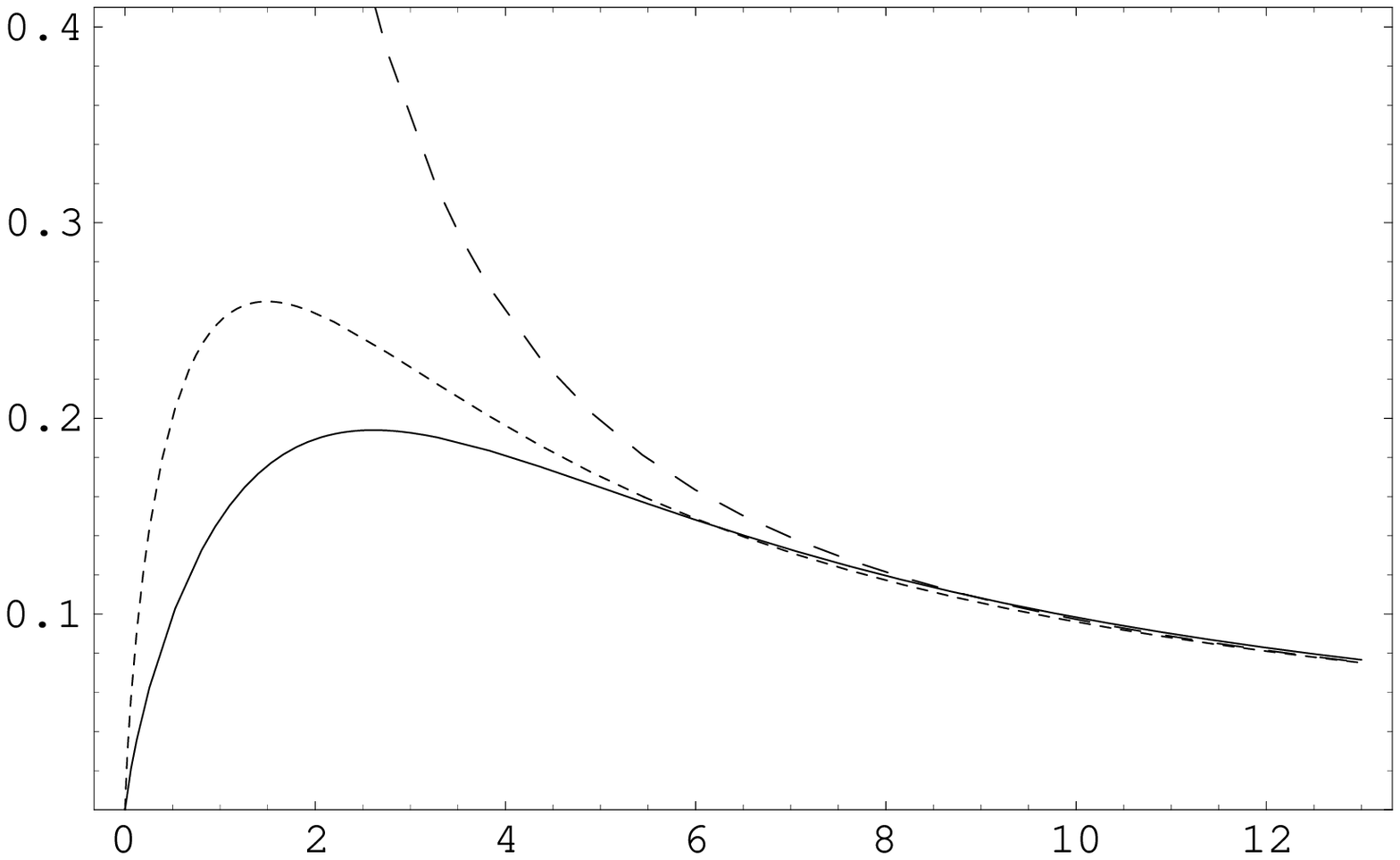,width=9cm}}
\put(4.5,0.0){$r/\varepsilon$}
\put(-1.0,4.4){$\sigma_{r\varphi}$}
\end{picture}
}
\caption{The components of stress:
(a)~$\sigma_{rr}$ and (b)~$\sigma_{\varphi\varphi}$  
are given in units of $\mu b_z/[2\pi(1-\nu)\varepsilon]$ for $\varphi=3\pi/2$
and  (c)~$\sigma_{r\varphi}$ is given in units of $\mu b_z\nu /[\pi(1-\nu)\varepsilon]$
for $\varphi=0$.  
with $(\beta+\gamma)(1-\nu)/\mu=2$ and $l=2\varepsilon$. 
The dashed curves represent the stresses in micropolar elasticity and 
the small dashed curves the stresses in strain gradient elasticity.}
\label{fig:T-edge-cc}
\end{figure}

Again, 
the stresses~(\ref{T_rr-cc})--(\ref{T_zz-cc}) are 
solutions of Eq.~(\ref{FE-NL}) in nonlocal micropolar elasticity. 
This nonlocal solution coincides with the solution given by
Povstenko and Matkovskii~\cite{PM97}.
In the limit $\varepsilon\rightarrow 0$,
we recover in (\ref{T_rr-cc})--(\ref{T_zz-cc}) 
the micropolar result given by~\cite{Kessel70,Nowacki74}.

Using~(\ref{CE2-inv}), we obtain the 
elastic micropolar distortion from the asymmetric force stress tensor
\begin{align}
\label{G_rr}
\gamma_{rr}&=-\frac{ b_x}{4\pi(1-\nu)}\, 
\frac{\sin\varphi}{r}\Big\{(1-2\nu)-\frac{4\varepsilon^2}{r^2}+2 K_2(r/\varepsilon)
+2\nu\, \frac{r}{\varepsilon}\, K_1(r/\varepsilon)\Big\}\nonumber\\
&\quad+\frac{(\beta+\gamma)b_x}{4\pi\mu}\, \frac{\sin\varphi}{r^3}
\Big\{2-\frac{r^2}{l^2-\varepsilon^2}
\big[K_2(r/l)-K_2(r/\varepsilon)\big]\Big\},\\
\label{G_pp}
\gamma_{\varphi\varphi}&=-\frac{b_x}{4\pi(1-\nu)}\, 
\frac{\sin\varphi}{r}\Big\{(1-2\nu)+\frac{4\varepsilon^2}{r^2}
-2 K_2(r/\varepsilon)-2(1-\nu)\,\frac{r}{\varepsilon}\, K_1(r/\varepsilon)\Big\}\nonumber\\
&\quad-\frac{(\beta+\gamma)b_x}{4\pi\mu}\, \frac{\sin\varphi}{r^3}
\Big\{2-\frac{r^2}{l^2-\varepsilon^2}
\big[K_2(r/l)-K_2(r/\varepsilon)\big]\Big\},\\
\label{G_rp}
\gamma_{r\varphi}&=\frac{b_x}{4\pi(1-\nu)}\, 
\frac{\cos\varphi}{r}\Big\{1-\frac{4\varepsilon^2}{r^2}+2 K_2(r/\varepsilon)\Big\}\\
&\quad-\frac{(\beta+\gamma)b_x}{4\pi\mu}\, \frac{\cos\varphi}{r^3}
\Big\{2-\frac{r^2}{l^2-\varepsilon^2}
\Big[K_2(r/l)-K_2(r/\varepsilon)
+\frac{\eta+\mu}{2\eta}\,
\Big(\frac{r}{l}\, K_1(r/l)-\frac{r}{\varepsilon}\, K_1(r/\varepsilon)\Big)\Big]\Big\},\nonumber\\
\label{G_pr}
\gamma_{\varphi r}&=\frac{b_x}{4\pi(1-\nu)}\, 
\frac{\cos\varphi}{r}\Big\{1-\frac{4\varepsilon^2}{r^2}+2 K_2(r/\varepsilon)\Big\}\\
&\quad-\frac{(\beta+\gamma)b_x}{4\pi\mu}\, \frac{\cos\varphi}{r^3}
\Big\{2-\frac{r^2}{l^2-\varepsilon^2}
\Big[K_2(r/l)-K_2(r/\varepsilon)
+\frac{\eta-\mu}{2\eta}\,\Big(\frac{r}{l}\, K_1(r/l)-\frac{r}{\varepsilon}\, K_1(r/\varepsilon)\Big)
\Big]\Big\},\nonumber
\end{align}
and
\begin{align}
\label{G_kk}
\gamma_{kk}=-\frac{\mu b_x(1-2\nu)}{2\pi(1-\nu)}\, 
\frac{\sin\varphi}{r}\Big\{1-\frac{r}{\varepsilon}\, K_1(r/\varepsilon)\Big\}.
\end{align}
It can be seen that the elastic micropolar strain $\gamma_{(ij)}$ is 
influenced by the
Cosserat constants $\beta$, $\gamma$ and $l$ and, therefore, it is different
from the elastic strain~(\ref{E_rr})--(\ref{E_pp})
of an edge dislocation in strain gradient 
elasticity.
Only the first parts of Eqs.~(\ref{G_rr})--(\ref{G_pr})
agree with the elastic strains~(\ref{E_rr})--(\ref{E_pp}) 
in strain gradient elasticity.
The other parts of Eqs.~(\ref{G_rr})--(\ref{G_pr})
are modified stresses due to the 
gradient micropolar elasticity. 
On the other hand, the micropolar dilatation~(\ref{G_kk}) has the same form as
the dilatation~(\ref{E_kk}) in strain gradient elasticity.
In the limit $\varepsilon\rightarrow 0$,
we recover in (\ref{G_rr})--(\ref{G_pr}) 
the micropolar result given by~\cite{Minagawa79}.

Using Eqs.~(\ref{mu-SFA}) and (\ref{SF2-ed}), we find for the micropolar
couple stresses
\begin{align}
\label{m_zr}
&\mu_{zr}=\frac{(\beta+\gamma)b_x}{2\pi}\,
\frac{\cos\varphi}{r^2}\Big\{1-\frac{1}{l^2-\varepsilon^2}
\Big[lr\, K_1(r/l)-\varepsilon r\, K_1(r/\varepsilon)
+r^2 \big(K_0(r/l)-K_0(r/\varepsilon)\big)\Big]\Big\},\\
&\mu_{z\varphi}=\frac{(\beta+\gamma)b_x}{2\pi}\,\frac{\sin\varphi}{r^2}
\Big\{1-\frac{1}{l^2-\varepsilon^2}
\Big[ lr\, K_1(r/l)-\varepsilon r\, K_1(r/\varepsilon)\Big]
\Big\},\\
\label{m_zp}
&\mu_{rz}=\frac{\beta-\gamma}{\beta+\gamma}\,\mu_{zr},\\
\label{m_pz}
&\mu_{\varphi z}=\frac{\beta-\gamma}{\beta+\gamma}\,\mu_{z\varphi}.
\end{align}
Due to the special choice of cylindrical coordinates the couple stresses 
(\ref{m_zr})--(\ref{m_pz})
have an artifical discontinuity at $r=0$. At $r=0$ the radial part of the 
couple stresses is finite and is multiplied by $\cos\varphi$ or $\sin\varphi$.   
In Cartesian coordinates the discontinuity disappears (see~\cite{LM04b}).
The couple stresses~(\ref{m_zr})--(\ref{m_pz}) 
are in agreement with the nonlocal couple stresses 
calculated by Povstenko and Matkovskii~\cite{PM97}.
In the limit $\varepsilon\rightarrow 0$,
we recover in (\ref{m_zr})--(\ref{m_pz}) 
the micropolar result given by~\cite{Kessel70}.

With~(\ref{CE2-inv}) we obtain from the micropolar couple stresses  
the corresponding micropolar bend-twist (wryness)
\begin{align}
\label{k-zr}
&\kappa_{zr}=\frac{b_x}{2\pi}\,
\frac{\cos\varphi}{r^2}\Big\{1-\frac{1}{l^2-\varepsilon^2}
\Big[lr\, K_1(r/l)-\varepsilon r\, K_1(r/\varepsilon)
+r^2\big(K_0(r/l)-K_0(r/\varepsilon)\big)\Big]\Big\},\\
&\kappa_{z\varphi}=\frac{b_x}{2\pi}\,\frac{\sin\varphi}{r^2}
\Big\{1-\frac{1}{l^2-\varepsilon^2}
\Big[lr\, K_1(r/l)-\varepsilon r\, K_1(r/\varepsilon)\Big]\Big\}.
\end{align}
The micropolar rotation can be calculated from the micropolar bend-twist.
For example, if we use
\begin{align}
\varphi_z=r\int\kappa_{z\varphi}\,\d\varphi,
\end{align}
we obtain
\begin{align}
\label{rot-z-ed}
\varphi_{z}=-\frac{b_x}{2\pi}\,\frac{\cos\varphi}{r}
\Big\{1-\frac{1}{l^2-\varepsilon^2}
\Big[lr\, K_1(r/l)-\varepsilon r\, K_1(r/\varepsilon)\Big]\Big\} .
\end{align} 
For $\varepsilon\rightarrow 0$, (\ref{rot-z-ed}) coincides with 
Nowacki's result~\cite{Nowacki86} calculated in the theory of micropolar elasticity.
The components of the force and couple stress, distortion and bend-twist tensors 
in Cartesian coordinates can be found in Ref.~\cite{LM04b}. 

\section{Conclusion}
In this paper, we have given an overview over static theories of gradient elasticity 
and nonlocal elasticity of Helmholtz type. 
We have discussed the theory of defects (dislocations, disclinations)
in (gradient) elasticity as well as (gradient) micropolar elasticity. 
Such theories are incompatible gradient theories because the elastic 
distortion quantities are not simple gradients of the displacement and 
the rotation.   
We have shown the equivalence 
between the nonsingular stresses of screw and edge dislocation in gradient 
elasticity and the corresponding nonlocal stresses. 
In addition, we have investigated the relationship between gradient micropolar elasticity
and nonlocal micropolar elasticity of Helmholtz type. 
For the static case, we have proven the equivalence of the force stresses and couple stresses 
calculated in gradient micropolar elasticity and nonlocal micropolar elasticity. 
We have considered straight screw and edge dislocation.
In order to fulfill the equilibrium conditions we have used the stress function
method. These stress functions are Green's function of partial differential equations
of higher order.
The calculated force and couple stresses do not possess singularities in
the dislocation core region. 
For the gradient theories we have calculated the elastic distortion and 
bend-twist tensors of screw and edge dislocations. 
Unlike the nonlocal theories, where these fields still possess singularities
at the dislocation line, the quantities calculated in the gradient 
theories are nonsingular.     
All fields calculated in the theories of gradient elasticity or gradient 
micropolar elasticity have the correct limits 
to classical elasticity or to micropolar elasticity.

\subsection*{Acknowledgement}
This work was supported by the European Network RTN ``DEFINO''
with contract number HPRN-CT-2002-00198.

\end{document}